\documentclass[aps,twocolumn,groupedaddress]{revtex4-1}
\usepackage{graphicx}

\usepackage{amsmath}

\begin{document}

\title{Models of Saturn's Interior Constructed with Accelerated Concentric
    Maclaurin Spheroid Method}

\author{B. Militzer}

\affiliation{Department of Earth and Planetary Science, University of
  California, Berkeley, CA 94720}
\affiliation{Department of Astronomy, University of
  California, Berkeley, CA 94720}

\author{S. Wahl}
\affiliation{Department of Earth and Planetary Science, University of
  California, Berkeley, CA 94720}

\author{W. B. Hubbard}
\affiliation{Lunar and Planetary Laboratory, University of Arizona,
    Tucson, AZ 85721}

\begin{abstract}
  The {\it Cassini} spacecraft's Grand Finale orbits provided a unique opportunity to
  probe Saturn's gravity field and interior structure. Doppler measurements
  \citep{Iess2019} yielded unexpectedly large values for the gravity harmonics $J_6$,
  $J_8$, and $J_{10}$ that cannot be matched with planetary interior models that
  assume uniform rotation. Instead we present a suite of models that assume the
  planet's interior rotates on cylinders, which allows us to match all the observed
  even gravity harmonics. For every interior model, the gravity field is calculated
  self-consistently with high precision using the Concentric Maclaurin Spheroid (CMS)
  method. We present an acceleration technique for this method, which drastically
  reduces the computational cost, allows us to efficiently optimize model parameters,
  map out allowed parameter regions with Monte Carlo sampling, and increases
  the precision of the calculated $J_{2n}$ gravity harmonics to match the error bars
  of the observations, which would be difficult without acceleration. Based on our
  models, Saturn is predicted to have a dense central core of $\sim$15--18 Earth
  masses and an additional 1.5--5 Earth masses of heavy elements in the envelope.
  Finally, we vary the rotation period in the planet's deep interior and
  determine the resulting oblateness, which we compare with the value from radio
  occultation measurements by the Voyager spacecraft. We predict a rotation period of
  10:33:34 h $\pm$ 55s, which is in agreement with recent estimates derived from ring
  seismology.
\end{abstract}



\maketitle

\section{Introduction}


Although Saturn's deep interior was not a primary target of the {\it Cassini}
spacecraft's 13-year mission monitoring the Saturnian system, the final phase of the
mission provided unprecedentedly  precise measurements of the planet's gravitational
field \citep{Iess2019}.  This phase, from April 23 to Sept.~15, 2017, culminated in
22 Grand Finale orbits, during which the {\it Cassini} spacecraft dived between the
planet and its innermost ring.  These measurements were contemporaneous with the
ongoing {\it Juno} mission, which is providing analogous measurements for Jupiter
\citep{Folkner2017a}. As a result of both studies, the measured gravity fields are
far more precise than ever before, warranting a closer look at the theory and
numerical techniques linking  the observed gravity to the interior density structure
of the planet.  Here we present models of Saturn's interior structure and interior
rotation rate, matched to the {\it Cassini} measurements, along with an acceleration
technique for the Concentric Maclaurin Spheroid (CMS) method \citep{Hubbard2013} for
calculating a self-consistent shape and gravity field.

Prior to {\it Cassini's} Grand Finale, the best determination of Saturn's gravity
field was from earlier flyby missions and from perturbations of the orbits of
Saturn's natural satellites in combination with the orbit of {\it Cassini} itself
\citep{Jacobson2006}. However, this yielded significant measurements of only the
first three even zonal harmonics of the field, $J_2$, $J_4$ and $J_6$.  By contrast,
X-band Doppler measurements during five of the 22 Grand Finale orbits produce a fit
with significant determination of even zonal harmonics up to $J_{12}$, as well as odd
zonal harmonics $J_3$ and $J_5$ \citep{Iess2019}. 

The distribution of mass within a planet depends on the equation of state of
hydrogen-helium mixtures at high pressures~\citep{MH13}, as well as the radial
distribution of heavier elements~\citep{Soubiran2016}. The interior
density distribution influences the observed structure of the gravity field through
deviations from spherical symmetry arising from rotation and tides. Thus, the
measured field can place constraints, albeit non-uniquely, on the internal structure
of the planet. For the rapidly rotating Jovian planets, such terms are primarily
determined by the balance between centrifugal and gravitational forces. In the
absence of internal dynamics, the density distribution and resulting gravity field
are axisymmetric and north-south symmetric, implying that only even zonal harmonics
$J_{2n}$ contribute to the gravitational potential.

\begin{figure}[b]
\includegraphics[width=8cm]{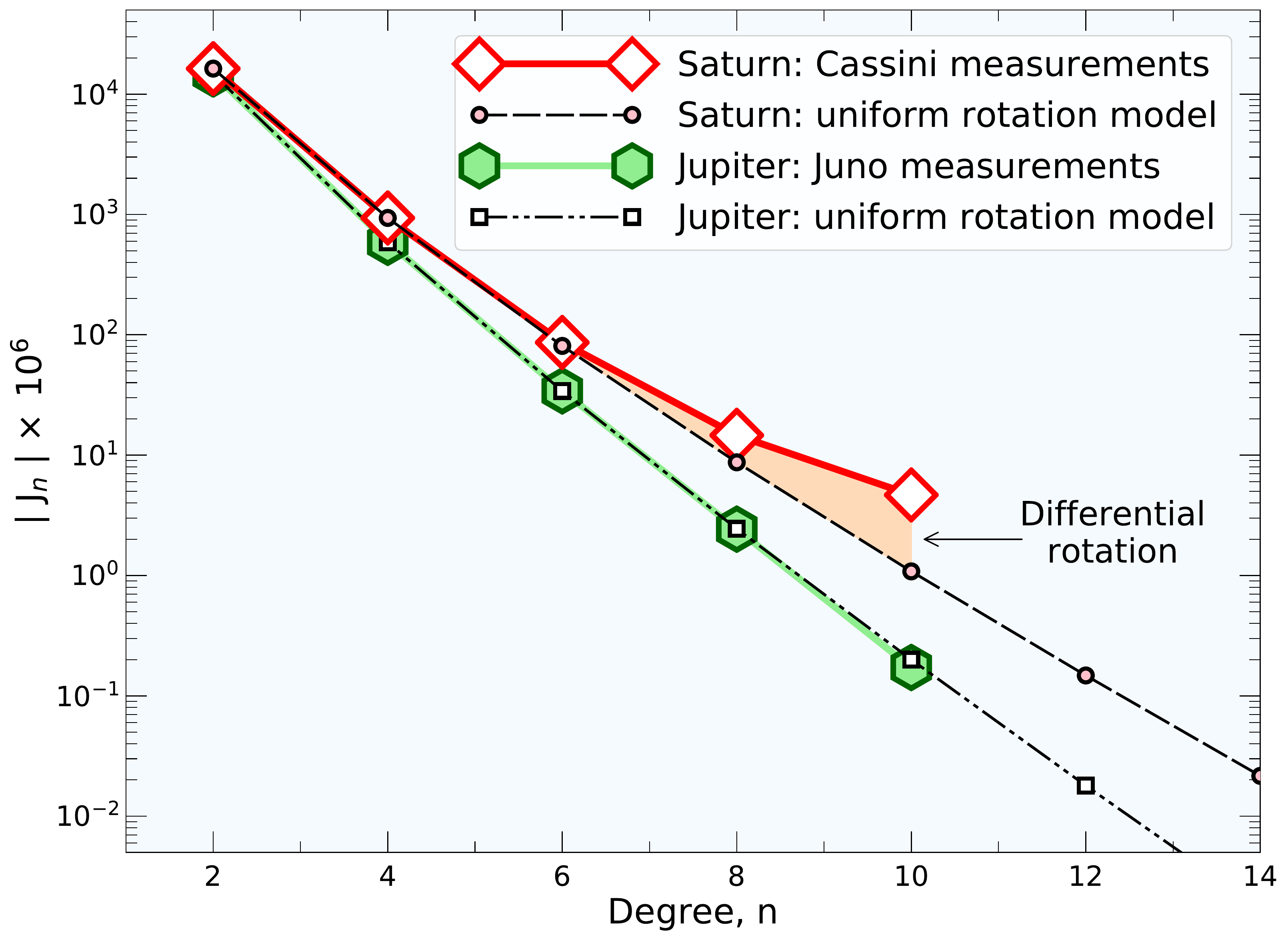}
\caption{Comparison of the gravity harmonics measured for Jupiter and
    Saturn with predictions from models assuming uniform rotation
  throughout the entire interior of both planets. The deviations are
  small for Jupiter while substantial discrepancies emerge for
  Saturn. This illustrates that the effects of differential rotation
  are much more important for Saturn.}
  \label{JupSatFig}
\end{figure}

If a planet in hydrostatic equilibrium rotates uniformly like a solid body, the
magnitudes of even zonal harmonics decay as $|J_{2n}|\sim q_{\rm rot}^n$, where
$q_{\rm rot}$ is the ratio of the centrifugal and gravity accelerations at the
equator. The $J_{2n}$ of Jupiter measured by {\it Juno} spacecraft are broadly
consistent with this relationship \citep{Folkner2017a}, meaning that it is possible
to find models with a uniform rotation rate that match the observed $J_{2n}$, at
least in the absence of other constraints, from the hydrogen-helium equation of state
and atmospheric composition.  However, Fig.~\ref{JupSatFig} illustrates how
the observed even moments $J_8$ and higher for Saturn deviate 
significantly from the expected relationship.  \citet{Iess2019} demonstrated that
these observations cannot be reproduced with models that assume uniform rotation, and
that deep differential rotation \citep{Hubbard1982} is required instead. In this
paper we expand upon the interpretation of \citet{Iess2019} and introduce new
analytical tools for high-precision gravity modeling.

\subsection{Differential Rotation}

Over many years prior to and including the duration of the {\it Cassini}
mission, optical tracking of clouds has revealed large-scale zonal wind currents
with respect to the average Saturn atmosphere, in particular a pronounced eastward
jet centered on the equator \citep{SL2000,GM2011}.  However, prior to the gravity
measurements discussed here, the data were insufficient to constrain the
depth of such zonal flows, and their effects were not considered in
previous modeling studies of Saturn's interior~\citep{HG13,N13}.  With the Grand
Finale gravity data, it becomes possible to test a model in which the cloud-level
zonal wind belts are mapped onto cylinders that extend to great depths. If the
zonal-wind velocity profile continues to depths of many scale heights, it will affect
the observed gravity field in two ways. First, it modifies the axisymmetric
gravitational field, and thus changes the even $J_{2n}$ from the values expected for
a uniformly rotating body with identical internal structure \citep{Hubbard1982}.
Second, to the extent that the velocity profile is not north-south symmetric, there
arises a corresponding asymmetry in the gravity field, manifesting itself in non-zero
odd $J_n$ \citep{Kaspi2013}. The values of $J_3$ and $J_5$ reported by
\citet{Iess2019} thus exhibit the north-south asymmetric component of the
differential rotation.

There are currently two basic methods for incorporating differential rotation into
gravity models. The first is to approximate the wind profile as rotation on
cylinders, which can be described using potential theory \citep{Hubbard1982} and can
therefore be integrated directly into the potential used in the CMS simulation
\citep{wisdom2016}. This method has the benefit of being fully self-consistent; the
dynamic contribution to the potential modifies the shape of the equipotential
surfaces, which feeds back into the calculated gravitational field. The downside is
that the wind profile must be constant on cylindrical surfaces and thus cannot decay
inward, as would be expected due to interactions with the magnetic field as hydrogen
becomes increasingly more conductive with increasing pressure~\citep{Cao2017}. For
instance, winds at high latitude could not be included in this method, because they
would correspond to cylinders extending all the way through the center of the planet.
Differential rotation on cylinders is also north-south-symmetric by definition, so
the odd $J_n$ are identically zero and cannot be modeled. The models presented in
this paper are subject to these limitations. 

The second method starts with a gravity solution assuming uniform
rotation, using CMS or a similar method, and then uses the thermal wind
equation~\citep{Kaspi2013,Galanti2017} or the gravitational thermal wind
equation~\citep{kong2013} to calculate a correction to the density and gravitational
moments. While this introduces additional approximations and does not produce a
self-consistent solution for the gravitational field, it allows for more flexible
wind fields, including cylinders of finite depth and flows with north-south
asymmetries. \citet{Iess2019} includes calculations in which the observed $J_n$ are
calculated with a decaying wind profile based on the observed cloud-level winds.

Nevertheless, the models with differential rotation on cylinders that do not decay
with depth are an important class of endmember models to consider for two reasons.
They fit all even gravity moments measured by the {\it Cassini} spacecraft and they
are fully self-consistent, which means that predictions for the core
mass, composition of the envelope and rotation profile will be obtained from just one
theory.  

\subsection{Interior Model Background}

Interior models of Saturn, like the ones presented here, have previously been fitted
to gravity data from {\it Voyager} \citep{ZG99,G99,SG04} and pre-Grand Finale {\it
Cassini} data \citep{HG13,N13}. In all cases they take into account a reduction of
helium mass fraction ($Y$) in the outer envelope arising from the immiscibility and
rainout of helium \citep{stevenson-astropj-77-ii}, although there are some
differences in the degree of rainout considered. The models differ primarily in the
material equations of state used, whether the heavy element concentrations ($Z$) are
homogeneous or inhomogeneous between the inner and outer envelope, and whether they
consider differential rotation. The range of predicted core masses decreased from
$\sim$10 -- 25 to $\sim$5 -- 20 Earth Masses when models were fitted to {\it
Galileo}-era and pre-Grand Finale {\it Cassini} gravity data \citep{Fortney2016}, and
some models considering inhomogeneous $Z$ had no central core at all \citep{HG13}.

One persistent issue for modelling Saturn's interior has been the uncertainty of the
planet's deep rotation rate, due to the near-perfect alignment of the magnetic field
dipole with the rotation axis. Given this uncertainty, we constructed ensembles of
models for four published rotation periods: 10:32:45 h \citep{Helled2015}, 10:39:22 h
\citep{desch1981}, 10:45:45 h \citep{Gurnett2005}, and 10:47:06 h
\citep{giampieri2006}. We also considered a very short rotation period of 10:30:00 h
in order to make the following calculation more robust. An independent constraint on
the rotation are measurements of the planet's degree of flattening (oblateness)
\citep{Lindal1985}. In Section \ref{sec:Ob}, we use this information to derive a new
estimate for Saturn's deep rotation period that is fully consistent with our interior
models, CMS method, and the {\it Voyager} oblateness measurements.

\section{Methods} \label{sec:method}

\subsection{Interior models}

\begin{figure}[ht!]
\includegraphics[width=4cm]{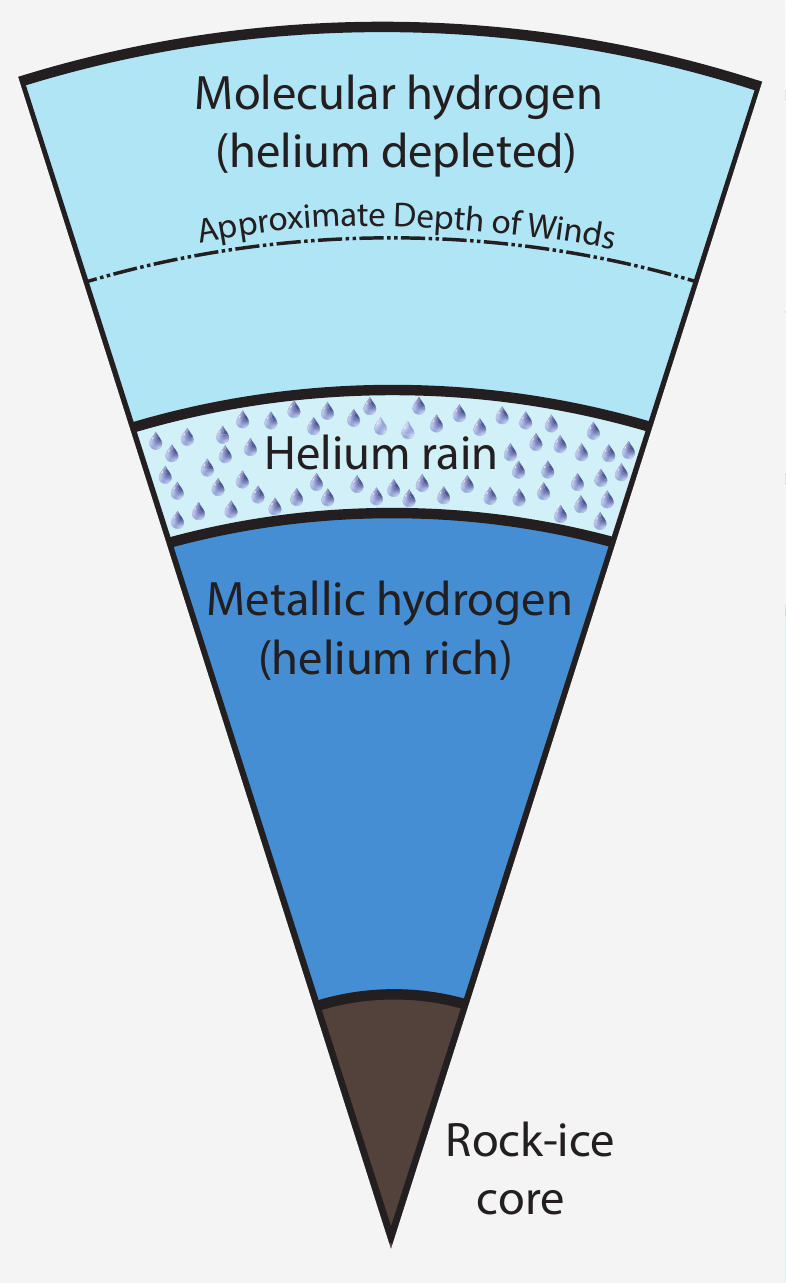}
  \caption{Illustration of the four-layer models for Saturn's interior that we
      constructed in this work. We assume an outer molecular and an inner metallic
      envelope, separated by a helium rain layer, with a dense core at
      the center of the planet.
\label{fig_saturn}}
\end{figure}

\begin{figure}[ht!]zo
\includegraphics[width=8cm]{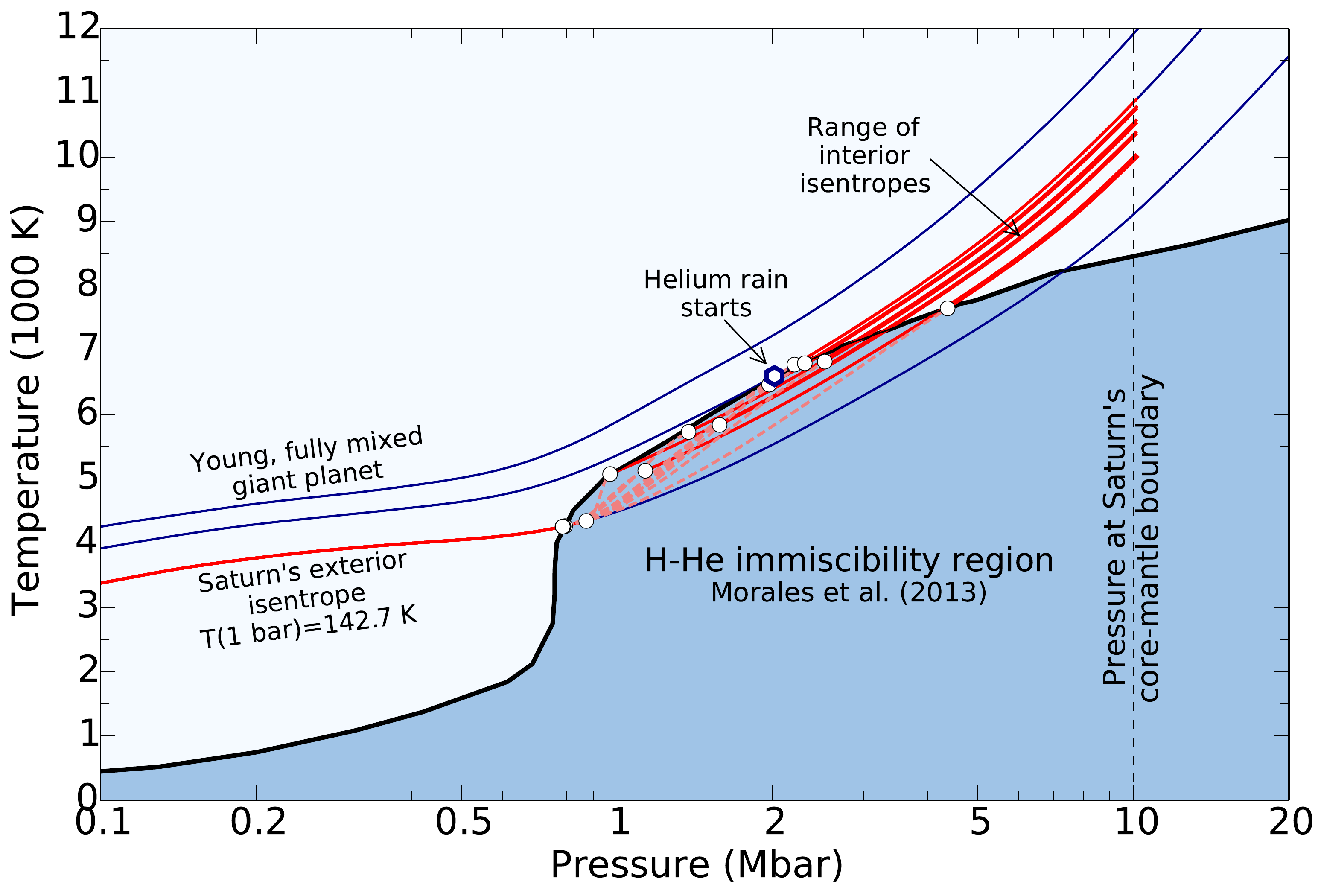}
  \caption{Temperature-pressure phase diagram of hydrogen-helium
    mixtures. The red lines show the interior models with
    $P$=10:39:22 h and $r_C$=0.2 in relation to the shaded region where
    the two fluids are predicted to become immiscible
    \citep{Morales2013}. The thin blue lines show various adiabats of
    an H-He mixture for a helium mass fraction of
    Y=0.245~\citep{Lodders03}. The circles mark the beginning and the
    end of the immiscibility regions assumed in different
    models. \label{fig_imm}}
\end{figure}

Since planets cool by convection, models are typically constructed
under the assumption that most regions in their interiors are adiabatic. However,
novel ideas based on double-diffusive convection have also been
considered~\citep{Leconte2013,Nettelmann2015}. One example of
non-adiabatic behavior occurs at high pressure, where hydrogen and helium are
predicted to become immiscible because hydrogen turns metallic while helium remains
an insulating fluid~\citep{stevenson-astropj-77-ii}, leading to a region
of helium rain. Following earlier work~\citep{Wahl2017a,Iess2019}, we assume
four-layer models with an outer molecular and an inner metallic
envelope, separated by a helium rain layer, along with a dense core at the center of
the planet, as illustrated in Fig.~\ref{fig_saturn}. In both envelope layers, an
adiabat consistent with {\it ab initio} simulations of hydrogen-helium
mixtures~\citep{Vo07,Militzer2013,MH13} is determined. Each adiabat is characterized
by an entropy, $S$, a helium mass fraction, $Y$, and a mass fraction of heavy
elements, $Z$. We adopt the phase diagram for hydrogen-helium mixtures as derived by
\citet{Morales2009}, and assume that helium rain occurs wherever the $P$-$T$
barotrope falls within the region of immiscibility in Fig.~\ref{fig_imm}.

We treat the helium rain layer as a smooth transition from the
parameters in the outer envelope ($S_{\rm mol}$, $Y_{\rm mol}$,
$Z_{\rm mol}$) to inner envelope ($S_{\rm met}$, $Y_{\rm met}$,
$Z_{\rm met}$) across a range of pressures $P_1$ to $P_2$, defined by
the intersections of the adiabat with the immiscibility curve. A
summary of our model parameters is given in Tab.~\ref{tab3}. A
collection of representative barotropes are shown in Fig.~\ref{fig_imm}.

Various core masses and radii are considered, but are not independent,
since the total mass of the core and envelope must match that of
Saturn. We first assumed fractional radius of 0.2 and later refined
the core radii by assuming either a terrestrial iron-silicate composition
(0.325:0.675) or a solar iron-silicate-water ice composition
(0.1625:0.3375:0.5). We find the fractional core radii of $r_C$=0.188
and 0.231 respectively to be consistent with these two
  compositions. We derived these core radii by adopting the additive
  volume rule for homogeneous mixtures in combination with the
  equations of state for iron, MgSiO$_3$ and water ice reported in
  \citet{Seager07} and \citet{WilsonMilitzer2014} that relied
  on experimental data and results from {\it ab initio} simulations.

For each set of model parameters, the CMS method finds a shape and
gravitational field for the planet consistent with a prescribed
rotation rate.

The distribution of helium across the rain layer is represented by a gradual
gradient with depth between $Y_{\rm mol}$ and $Y_{\rm met}$. Thus a value of
$Y_{\rm mol}$, up to the solar helium fraction $Y$=0.274 \citep{Lodders03}, is
considered and a consistent $Y_{\rm met}$ above the solar fraction is
determined such that total, planet-wide helium mass fraction is conserved. The
entropy of the outer envelope adiabat $S_{\rm mol}$ is chosen to be consistent
with the observed temperature 142.7 K at 1 bar ~\citep{Lindal1981}.

\subsection{Concentric Maclaurin Spheroid Method}

The literature on the problem of the shape and gravitational potential
of a liquid planet in hydrostatic equilibrium (also referred to as the
theory of figures, TOF) extends back centuries \citet{jeans1919}. Most
geophysical implementations of TOF use a perturbation approach, by
finding the response, to various orders, to a small perturbation of
the potential from spherical symmetry. For a discussion of
perturbation TOF, see \citet{ZT1978}.

\citet{Hubbard2012} developed a non-perturbative numerical method,
based on potential theory~\citep{tassoul2015}, for calculating the
self-consistent shape and gravitational field of a constant density,
rotating fluid body to high precision.  This method was generalized to
approximate a barotropic pressure-density relationship, discretizing
the interior into a series of concentric constant-density (Maclaurin)
spheroids (CMS) by \citet{Hubbard2013}. The spheroids comprise
constant-potential level surfaces, deformed in two dimensions for
permanent rotation about a fixed axis, and in three dimensions if a
tidal potential is included \citep{Wahl2017b}. Thus, the surface of
every spheroid is a surface of constant potential, density, pressure,
temperature, and composition.  The CMS method is non-perturbative and
thus more general than methods that approximate the level surfaces as
perturbed ellipsoids.  The CMS method has been benchmarked against
an independent, non-perturbative numerical method \citep{wisdom2016}.

In this paper, we introduce an accelerated version of the CMS method, in which the
shape of a subset of spheroids is calculated explicitly, with the shape of most
spheroids obtained through interpolation of the radius. As we will show, this leads
to a much more efficient algorithm for the same level of precision of the predicted
gravity field. The acceleration technique enables us to construct ensembles of
Saturn's interior models with Monte Carlo sampling and to perform proof-of-principles
CMS calculations with a large number of layers ($N_L$) $\sim 10^5$. Both would not have been
feasible without acceleration of the method.

As noted by \citet{Debras2018}, while a model with a given number of spheroids
generates an external gravity potential to a numerical precision of at least
$10^{-12}$ (much better than {\it Juno} or {\it Cassini} measurement precision), the
precision to which it approximates the smooth $\rho(P)$ barotrope is limited by
the number of layers. This leads to an $N_L$-sensitivity of the generated gravity
potential that is larger than the uncertainty in the measured potential, as initially
quantified by \citet{wisdom2016}. The acceleration to the CMS method helps us rectify
any uncertainty from discretization, allowing a much smoother discretization of the
barotrope while the more computationally expensive part of the method is kept to a
manageable number of layers. 

\subsection{ Self-consistent Shape and Gravity with CMS }

The CMS technique, based on potential theory, allows one to describe the interior
of planets under the assumption of hydrostatic equilibrium. Baroclinic effects are
excluded from consideration, which implies that the temperature of a fluid parcel is
only a function of its pressure, $T(P)$. While this is well justified in the deep
interior, it is more of an approximation at the 1 bar level when we relate the
temperature of fluid parcels near the equator with those in the less irradiated polar
regions. Under this assumption, we combine $T(P)$ with a realistic equation
of state, $\rho=\rho(P,T)$, of a mixture of hydrogen, helium, and a small amount of
heavier elements in order to establish a barotrope, a unique density-pressure
relation $\rho(P)=\rho(P,T(P))$. This assumes knowledge of the composition as a
function of pressure.

In hydrostatic equilibrium, the pressure, $P$, the mass density,
$\rho$, and the total potential, $U$, at any point in the planet's
interior are related by
\begin{equation}
  \begin{aligned} \nabla P = \rho \; \nabla U.
  \end{aligned}
  \label{eq:hydrostatic}
\end{equation}
The sign of the potentials is chosen such that forces are given by
$F = + \nabla U$. In the co-rotating frame of the planet, the total
potential, $U$, includes contribution from the self-gravity, $V$, and
the centrifugal term, $Q$,
\begin{equation} 
U = V + Q,
\end{equation} 
which we discuss in detail in the two following sections. 

For a planet with a uniform rotation rate, it is convenient to describe the relative
strength of of the rotational perturbation in terms of the parameter
\begin{equation}
  q_{\rm rot} = \frac{\omega^2 a^3}{GM}, \label{eq:qrot}
\end{equation}
where $\omega$ is the rotation rate, $G$ is the universal
gravitational constant, and $M$ and $a$ are the mass and equatorial
radius of the planet.  Since CMS theory is non-perturbative, in
principle the results are valid to all powers of $q_{\rm rot}$.  

It follows that the pressure, density and potential can be expressed
in dimensionless, planetary units (PU):
\begin{equation}
  \begin{aligned}
    P_{\rm pu}   \equiv& \, \frac{a^4}{GM^2} \, P \;,\\
    \rho_{\rm pu}\equiv& \, \frac{a^3}{M} \, \rho \;,\;\; {\rm and}\\
    U_{\rm pu}   \equiv& \, \frac{a}{GM} \, U. \\
  \end{aligned}
  \label{eq:planetary_units}
\end{equation}

We label the $N_L$ spheroids with the indices $i=0,1,2,\dots,N_L-1$,
with $i = 0$ corresponding to the outermost spheroid and $i=N_L-1$
corresponding to the innermost spheroid. All models presented here
are symmetric with respect to the axis of rotation. We neglect any
non-axisymetric contributions to potential, such as tidal
perturbation by a satellite~\citep{WHM16a,WHM16b}. So the shape of
every spheroid $i$ can be described by a function $r_i(\mu)$ where
$r_i$ is the distance from the planet's center and $\mu=\cos(\theta)$
is a function of the polar angle, $\theta$. We assume throughout its
interior, the planet is north-south symmetric, which implies,
$r_i(\mu)=r_i(-\mu)$.

\begin{figure}[ht!]
\includegraphics[width=8cm]{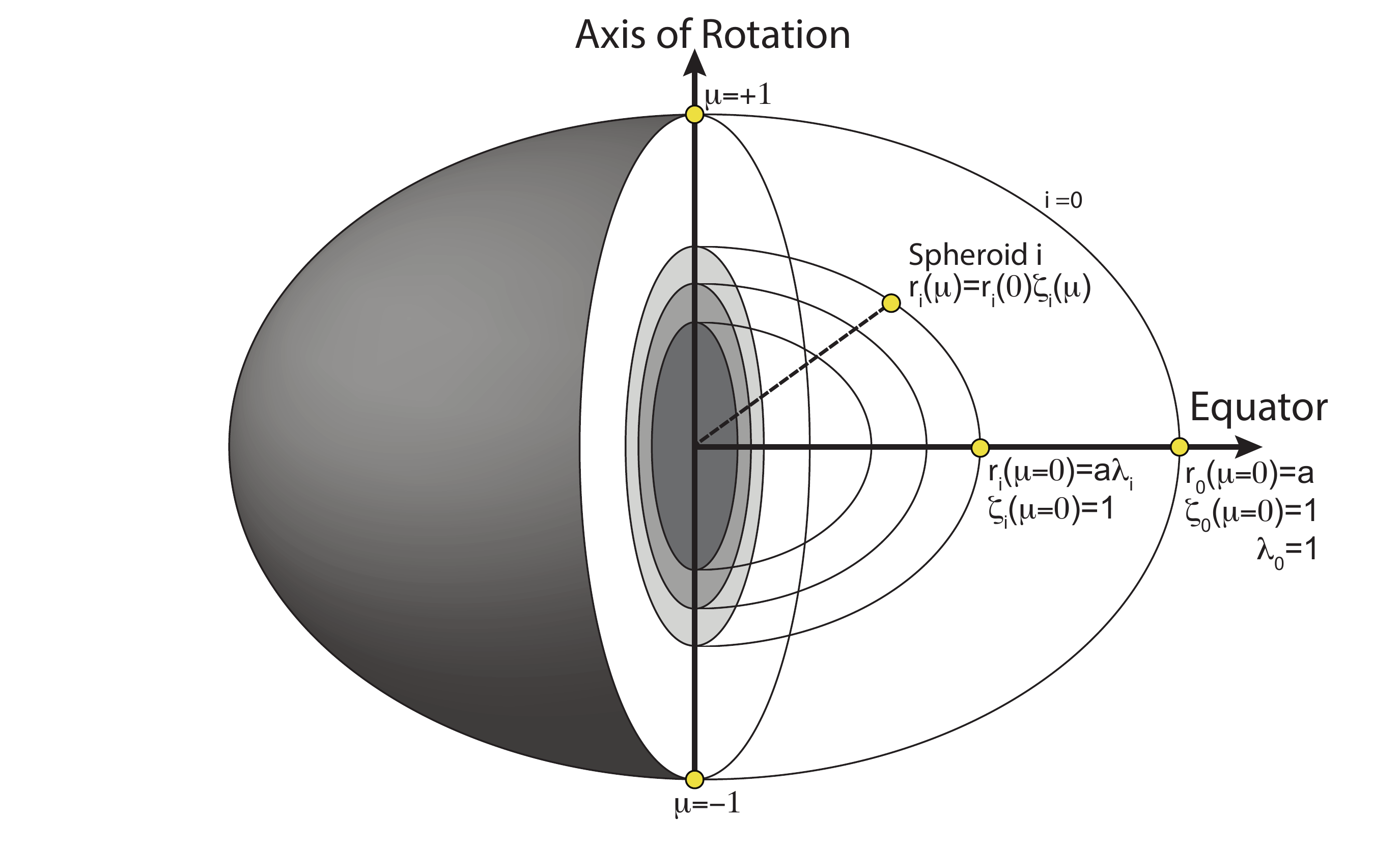}
  \caption{ Illustration of the CMS method and variable definitions.\label{fig_CMS} }
\end{figure}

It is convenient to introduce a normalized shape function,
\begin{equation}
    \zeta_i(\mu) \; \equiv \; \frac{r_i(\mu)} {r_i(0)} \; \leq \; 1\;
  \label{eq:shape}
\end{equation}
where $r_i(0)$ is equatorial radius of $i$th spheroid. $\zeta_i(\mu)$
will approach unity for non-rotating planets.  Furthermore, we define
$\lambda_i \equiv r_i(0) / r_0(0) $ to be the ratio of the equatorial radius of
the $i$th to the outermost spheroid. Note that
$r_0(0) \equiv a$. These choices are illustrated in
Fig.~\ref{fig_CMS}.

Hydrostatic equilibrium requires that the density increases
monotonically with depth and thus with spheroid index $i$. We can
define $\delta_i$ to be the density difference between two adjacent
spheroids,
\begin{equation}
  \delta_i = \begin{cases}
    \rho_i - \rho_{i-1}, & i>0 \\
    \rho_0, & i=0.\\
  \end{cases}
  \label{eq:density_increment}
\end{equation}
This parameterization of density has the added benefit of naturally handling
discontinuities in $\rho$, as would be expected for compositionally distinct
layers.

We represent the shape functions, $\zeta_i(\mu)$, on a grid of $N_\mu$
points, $\mu_m$, such that $\zeta_{im} \equiv \zeta_i(\mu_m)$.  The
CMS method refines the shape functions through an iterative procedure
until the potential on every spheroid surface is constant and the
equation of hydrostatic equilibrium is satisfied
(Eq.~\ref{eq:hydrostatic}).  In the current implementation, we keep
equatorial radii of every spheroid fixed, $r_i(0)=\lambda_i a$, while
the remaining spheroid points are adjusted until a self-consistent
solution has been found.

We start the iterations with all spheroids to be perfect spheres and
thus initialize all normalized shape functions to unity,
$\zeta_{im}=1$. A given set of spheroids defines a mass distribution
and thus a gravity field. We can define a function $U_i(\zeta,\mu)$ to
calculate the total potential on the surface of spheroid $i$. The
spheroid shape has converged if $U_i(\zeta,\mu)$ is the same for all
$\mu$. However, at the beginning there will always be significant
deviations that we can encapsulate in a function,
\begin{equation}
  f_{im}(\zeta_{im}) \equiv U_i(\zeta_{im},\mu_m) - U_i(1,0),
\end{equation}
that compares the potential at $\zeta_{im}$ and $\mu_m$ with that of
reference point on the equator of spheroid $i$. We compute the
derivative $f'_{im}(\zeta_{im}) = d f_{im}(\zeta_{im}) / d \zeta_{im}$
analytically and employ a single Newton step to derive an improved
value for $\zeta_{im}$ from
\begin{equation}
  \zeta_{im}^{\rm (new)} = \zeta_{im} - \frac{f_{im}(\zeta_{im})}{f'_{im}(\zeta_{im})}\;.
  \label{Newton}
\end{equation}

Once the points on all spheroids have been updated, we recalculate the
zonal gravitational moments, $J_{n}$, in order to obtain an updated
gravity field, $U_i$. Assuming hydrostatic equilibrium
(Eq.~\ref{eq:hydrostatic}), we successively update the pressure on
every spheroid
\begin{equation}
    P_{i}^{\rm (new)} = P_{i-1}^{\rm (new)} + \rho_{i-1} \, (U_{i}-U_{i-1}).
    \label{P_update}
\end{equation}
starting from $P_0$ that we keep fixed at 0.1 bar. This value is
consistent with the observed gravity harmonics that were normalized to
an equatorial radius of $a = 60330$ km~\citep{Iess2019}.

Next we update the density of every spheroid,
\begin{equation}
    \rho_{i}^{\rm (new)} = \rho( \, (P_{i+1}+P_i)/2 \, ),
    \label{eq:rhodesired}
\end{equation}
by evaluating the prescribed barotrope function,
$\rho(P)$, for the average of the pressure at the upper
and the lower boundaries of a particular spheroid.

After every improvement of the spheroid shapes, $\zeta_{im}$, an
update step for the gravity harmonics, the potential, pressure, and
spheroid densities follows. These two steps are repeated until all of
the moments, $J_{n}$, have converged such that the difference between
successive iterations falls below a specified tolerance. Occasionally,
we find the convergence of the algorithm to be slow if the shapes
oscillate back and forth between two states. We detect such events and
bypass them by inserting a {\it regula falsi} step.

It is also necessary to have at least one free parameter for a subset
of the layers in order to obtain the correct total mass of the CMS
model. In our implementation we modify the mass of the central core
to achieve this balance.

\subsection{Gravitational Potential}
\newcommand{\sphint}{\int_{-1}^1 d\mu' \int_{0}^{2\pi} d\phi' \int_{r' > r} dr'}
\newcommand{\sphshort}{\int_{\tau} d\tau}
\newcommand{\muint}{\int\limits_{-1}^{+1} d\mu}
\newcommand{\phiint}{\int_{0}^{2\pi} d\phi'}
\newcommand{\xiint}{\int_{b/a}^{\xi} d\xi'}
\newcommand{\til}{\widetilde}

The gravitational potential at a vector coordinate, $\mathbf{r}$, due to an arbitrary mass
distribution is given by 
\begin{equation} \label{eq:basicV}
    V(\mathbf{r}) = G \int d^3\mathbf{r'}
    \frac{\rho(\mathbf{r'})}{\left|\mathbf{r}-\mathbf{r'}\right|}.
\end{equation} 
In the case of an axisymmetric mass distribution with the center of mass at the
origin, the potential can be expanded in the following form~\citep{ZT1978},
\begin{eqnarray}
  V(r,\mu) &=& 
               \frac{G}{r} \sum_{n=0}^{\infty} P_n(\mu) \int \,d\tau'\,
               \rho(r') \, P_n(\mu') \, \left(\frac{r'}{r}\right)^{n}
               \label{eq:general_expansion}\\
           &=&
               \frac{GM}{r} \left[ 1 - \sum_{n=1}^{\infty} \left(a/r\right)^{2n} J_{2n}P_{2n}(\mu) \right]
               \quad.
\end{eqnarray}
where $d\tau' = r'^2 dr' \, d\mu' \, d\phi'$. $P_n$ are the Legendre polynomials of order $n$. The gravity harmonics are given by
\begin{equation}
  J_n = - \frac{2 \pi}{M a^n} \int\limits_{-1}^{+1} d \mu \int\limits_0^{r_{\rm max}(\mu)} \!\!\!\! dr \,\, r^{n+2} \,\, P_n(\mu) \,\, \rho(r,\mu)
  \label{standard_J}\quad.
\end{equation}
$J_0$ represents the integral over all mass and has been normalized to
equal $-1$ by convention.

Following \citet{Hubbard2013}, the self-gravity contribution to the
potential is found by expanding Eq.~(\ref{eq:general_expansion}) in
terms of the interior zonal harmonics, $J_{i,n}$, and the external
zonal harmonics, $J_{i,n}'$ and $J_{i,n}''$, for every spheroid $i$
and order $n$. At the surface of the planet, the observable zonal
harmonic is the sum of the moment from every spheroid.

For convenience, the harmonics are normalized by the equatorial radius
of the corresponding spheroid
\begin{equation}
  \til{J}_{i,n} \equiv \frac{J_{i,n}}{\lambda_i^n} \;\; {\rm and} \;\;
  \til{J}'_{i,n} \equiv J'_{i,n}\lambda_i^{(n+1)}. 
  \label{eq:normalization}
\end{equation}
Following the derivation in \citet{Hubbard2013}, we find the
normalized interior harmonics
\begin{equation}
    \til{J}_{i,n} = -\frac{1}{n+3} \, \frac{2 \pi }{M} \, \delta_i \lambda_i^3 \, \muint\ \, P_n(\mu) \, \zeta_i(\mu)^{(n+3)}
    \label{eq:harmonics}
\end{equation}
and the exterior harmonics
\begin{equation}
   \til{J}'_{i,n} = - \frac{1}{2-n} \, \frac{2 \pi}{M} \, \delta_i
   \lambda_i^3 \, \muint \, P_n(\mu) \, \zeta_i(\mu)^{(2-n)}
   \label{eq:harmonics_prime}
\end{equation}
with a special case for $n=2$
\begin{equation}
  \til{J}'_{i,n} = - \frac{2 \pi}{M} \, \delta_i \lambda_i^3
    \muint \, P_n(\mu) \, \log(\zeta_i) 
    \label{eq:harmonics_prime_n=2}
  \end{equation}
and
\begin{equation}
  J''_{i,0} = \frac{2\pi\delta_i a^3}{3 M},
  \label{eq:J_prime_prime}
\end{equation}
where $M$ is the total mass of the planet given by
\begin{equation}
M = \frac{2 \pi}{3} \sum_{i=0}^{N-1} \, \delta_i \, \lambda_i^3 \muint \, \zeta_i(\mu)^3
\end{equation}

With this description of the planet's self-gravity in terms of $J_{i,n}$, $J_{i,n}'$
and  $J_{i,n}''$, the expansion of of Eq. \ref{eq:general_expansion} for a point on
surface $i$ yields
\begin{eqnarray}
  \label{eq:Vj}
  V_i(\zeta_i,\mu) &=&
  -\frac{1}{\zeta_i \lambda_i} \left[ \sum^{N-1}_{j=i} \sum^\infty_{n=0} 
      \til{J}_{j,n} \left( \frac{\lambda_j}{\lambda_i \zeta_i} \right)^n P_{n}(\mu)
     \right.\\
     \nonumber
     &&+ 
     \left.
     \sum^{i-1}_{j=0}
    \sum^\infty_{n=0} \til{J}'_{j,n} \left( \frac{\lambda_i \zeta_i}{\lambda_j} \right)^{n+1}
    P_n(\mu)  
    + \sum^{i-1}_{j=0}
    J''_{j,0} \lambda_i^3 \zeta_i^3 \right].
       \end{eqnarray}
The gravitation potential on the equator of the outermost spheroid is given by 
\begin{equation}
  V_{i=0}(1,0) = 
  - \sum^\infty_{n=0} P_{n}(0)  J_{n}, 
\end{equation}
where
\begin{equation}
            J_{n} = \sum^{N-1}_{i=0} \lambda_i^n \til{J}_{i,n}
\end{equation}
are the standard zonal gravity harmonics of the observable surface
field in Eq.~\ref{standard_J}. 
In practical application of the CMS
method, one finds that results converge rapidly with increasing
polynomial order, $n$. So we typically terminate the sum over $n$ at
16 or 32.

\subsection{Centrifugal Potential}

\begin{figure}[ht!]
\includegraphics[width=8cm]{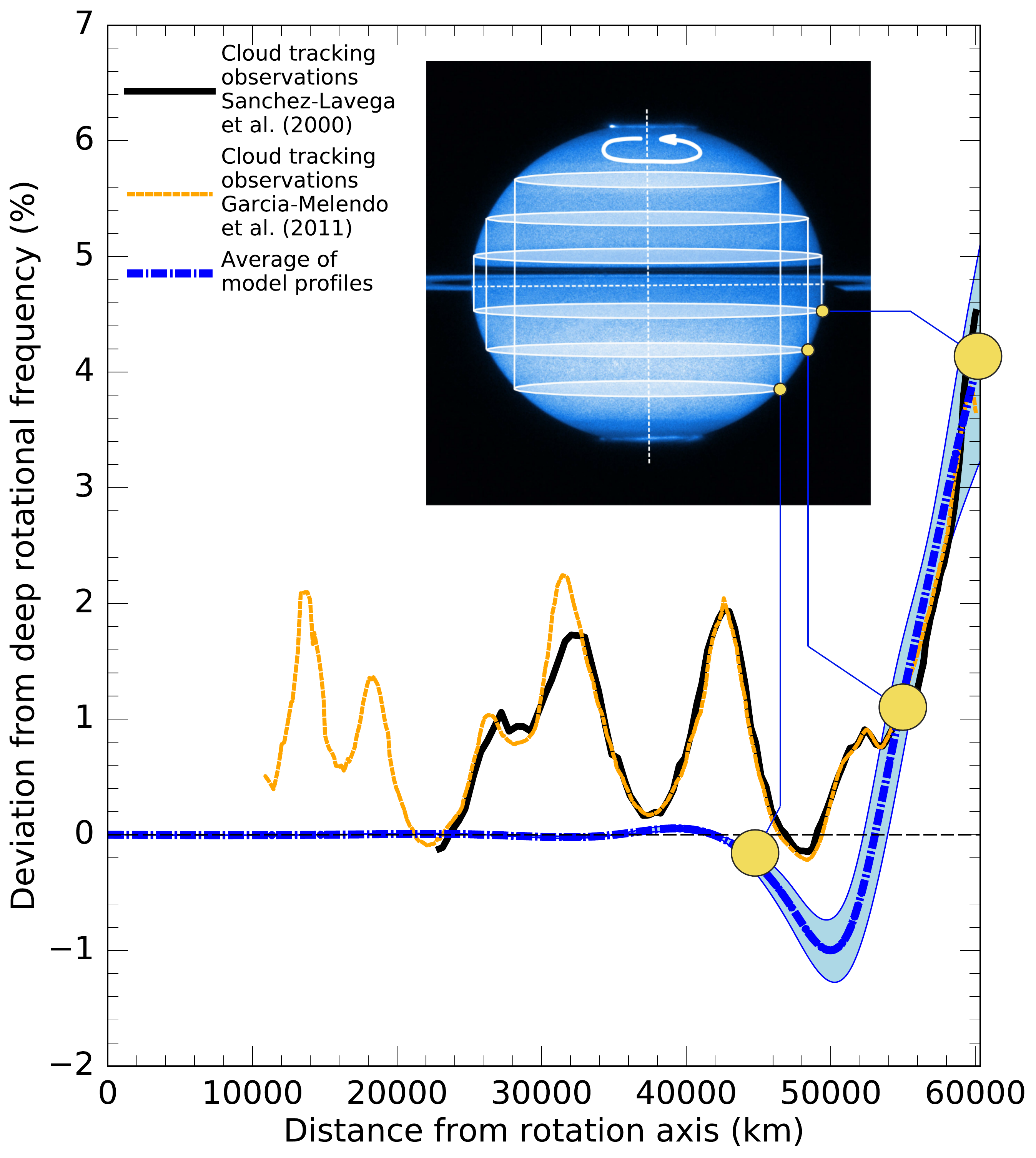}
\caption{Average of the rotation profiles in our suite of Saturn
  interior models that match the observed even gravity harmonics. It
  shows that differential rotation must be several thousands of
  kilometers deep. Our models reproduces the Eastward equatorial jet
  that rotates about 4\% faster than the deep interior. The inset shows an
  illustration of the cylinders. The rotation frequencies inferred by
  tracking the clouds in Saturn's visible
  atmosphere~\citep{SL2000,GM2011} are shown for comparison.}
\label{DR}
\end{figure}

We assume potential theory throughout this work and we are thus
restricted to studying two cases: uniform rotation ($\omega$ =
constant) and differential rotation on cylinders where the angular
frequency, $\omega(l)$, is solely a function of the distance from the
rotation axis, $l$. An illustration is shown in
Fig.~\ref{DR}. Everywhere the centrifugal force,
$\vec{F} = l\omega^2 \vec{e}_l$, is perpendicular to the axis of
rotation, which we assume to be the $z$ axis. In potential theory,
this force is represented by the centrifugal potential,
\begin{equation}
Q = \int_0^l dl' \; l' \; \omega(l')^2
\end{equation} 
If $\omega$ is constant, one recovers the usual term
$Q(l) = \frac{1}{2} l^2 \omega^2$. It is not possible to give the
cylinders a finite depth, $H$, within potential theory. Calculations
with finite $H$ can be performed with the thermal wind
equation~\citep{KG16,Kaspi2017a,Kaspi2018} or the gravitational
thermal wind equation~\citep{kong2013}. If one wanted to give the
cylinders a finite depth or introduce any other $z$ dependence,
$\omega(l,z)$, one would inevitably introduce spurious force terms
parallel to the $z$ direction because the derivative
$\partial Q / \partial z$ is no longer zero. This force would not be
consistent with the assumption that the centrifugal force should be
perpendicular to the axis of rotation~\citep{tassoul2015}. Therefore,
the cylinders in our calculations penetrate through the 
equatorial plane of the
planet. As we will later see, this allows us to reproduce the
observed winds in the equatorial regions but not those at higher
latitudes, because they would involve very deep cylinders with too much
mass.

Most simply, one can represent the angular frequency by an expansion
in even powers of $l$,
\begin{equation}
\omega(l) = \omega_0 \; + \; c_2 l^2 \; + \; c_4 l^4 + \; c_6 l^6 + \; c_8 l^8
\; + \; \ldots ,
\label{exp}
\end{equation} 
where $\omega_0$ is the rotation rate in the deep interior and the expansion
coefficients, $c_{2i}$, present the differential part. These
coefficients need to be optimized jointly with the parameters of
our interior model in order to reproduce the gravity coefficients that
were measured by the {\it Cassini} spacecraft. While the expansion in
Eq.~\ref{exp} may be convenient for analytical work, we found this
functional form to be impractical for numerical optimizations. If one
changes one coefficient in the expansion, rotation of all fluid
parcels is affected. Changing the rotation rate in a
small interval of $l$, requires changing several coefficients in a coordinated fashion. Such inter-dependencies are detrimental for the
efficiency of any optimization algorithm. We therefore represent the
angular frequency, $\omega(l)$, from $l=0 \ldots 1$ by a spline
function with a fixed number of knots, $l_k$, on which we adjust the
frequency $\omega(l_k)$. In this formulation, a change of
$\omega(l_k)$ will only affect fluid parcels between $l_{k-1}$ and
$l_{k+1}$, which greatly simplifies the optimization.

We obtained good results with 11 and 21 knots. Furthermore, in our
Monte Carlo (MC) calculations and simplex optimizations, we observed
that
the angular frequency $\omega(l_k)$ for radii interior to $l_k < 0.7$
never deviated from $\omega_0$, presumably because the associated
cylinders were so deep and involved too much mass. Based on these
observations, we exclude the $\omega(l_k)$ values for small $l$ from
the optimization and set $\omega(l_k<0.7) = \omega_0$ instead.

\subsection{Acceleration  of the CMS method}
 
Among numerical methods to solve partial differential equations (PDE),
one distinguishes between finite difference and finite element
techniques~\citep{Morton2005}. In the former approach, one
approximates the derivatives in the PDE by computing differences
between two adjacent points on the integration domain. In the more
sophisticated finite element approach, one also considers the
properties of the interior of every integration interval. This
typically enables one to derive a more accurate solution than is
possible with finite difference approaches, when the two methods are
compared for the same grid resolution.

The acceleration of the CMS method, that we will now introduce, is
comparable to switching from the finite difference to a finite element
approach. The goal is to reduce the primary discretization error of
the CMS methods that arises from the approximation that the density
changes in a step-wise fashion from one spheroid to the next. The
acceleration becomes possible because each CMS iteration has two parts that
have very different computational costs. The expensive part
(Eq.~\ref{Newton}) involves updating the shape of every spheroid
represented by the variables $\zeta_{jm}$ for a given gravity
field. In the second, comparatively cheap step, one updates the
interior and exterior gravity harmonics in
Eqs.~\ref{eq:harmonics}-\ref{eq:J_prime_prime} for the current
spheroid geometry. As it turns out, the accuracy of the computed
gravity harmonics depends sensitively on the number of spheroids,
$N_L$, which determines how precisely the smooth density profile in the
planet's interior is approximated by the step-wise representation of
the nested constant-density spheroids.

The core idea behind the acceleration is to only compute the
spheroid shape explicitly at every $n_{\rm int}$ layers. For the
$(n_{\rm int}-1)$ layers in between, we interpolate the shape
functions $\zeta_{im}$ as a function of $\lambda_i$ at constant
$\mu_m$. This $\zeta_{im}$ update is the most expensive part of the CMS
calculation and scales like $N_L \times N_\mu$ while the other
parts of the calculation all scale like $N_L$. Therefore, we evaluate the other parts
of the calculation over the entire set of $N_L$ spheroids as before. The
cost of the spline interpolation is negligible compared to the
explicit updates of the $\zeta_{im}$ points according to
Eq.~\ref{Newton}. The inner and outermost spheroids are always updated
explicitly to avoid extrapolations.

Instead of updating $\zeta_{im}$ for  $N_L$ layers, we only need to update
$ N_L/n_{\rm int}$ layers
\footnote{To keep the
following analysis simple, we write $N_L/n_{\rm int}$ for the number
of layers that we treat explicitly while it is in fact
$(N_L-1)/n_{\rm int}+1$.}. 
The reduction in computational cost can be reinvested into
increasing the total number of layers. As we will show, the accuracy
of an accelerated CMS computation with a total layer number of
$N_L^{\rm acc} = N_L^{\rm org} \times n_{\rm int}$ will be much higher
than that of the original calculation of $N_L^{\rm org}$ layers, while
both have comparable computational cost.  The computation of all
gravity harmonics is be performed with all $N_L^{\rm acc}$ layers,
which significantly improves the accuracy compared with the original
calculations with $N_L^{\rm org}$ layers.

In order to analyze the accuracy and the performance of our
acceleration technique, we constructed a representative model for the
interior structure of Saturn. For this analysis, we assume uniform
rotation and performed calculations for a variety of layer numbers
with interpolation parameter, $n_{\rm int}$, ranging from 2 to
128. The results of the original method without acceleration are
recovered for $n_{\rm int}=1$. The resulting gravity coefficients that
were computed with and without acceleration are compared in
Tabs.~\ref{tab1} and \ref{tab2} for different layer numbers. One finds
that all gravity harmonics converge smoothly as a function of layer number, which
allows one to extrapolate to $N_L \to \infty$. We infer $J_n(\infty)$
by employing the following semi-linear fit function:
\begin{equation}
\log \left|\Delta J_n(N_L) \right| \equiv \log \left| J_n(N_L) - J_n(\infty) \right| = A - B \log[N_L]
\end{equation}
For every gravity coefficient, $J_n$, we adjust the fit parameter
$J_n(\infty)$ and derive the linear fit coefficients $A$ and $B$ until
we have obtained the best possible match to the $J_n(N_L)$ data
set. The extrapolated values, $J_n(\infty)$, are included in
Tab.~\ref{tab2}.

\begin{figure}[ht!]
\includegraphics[width=8cm]{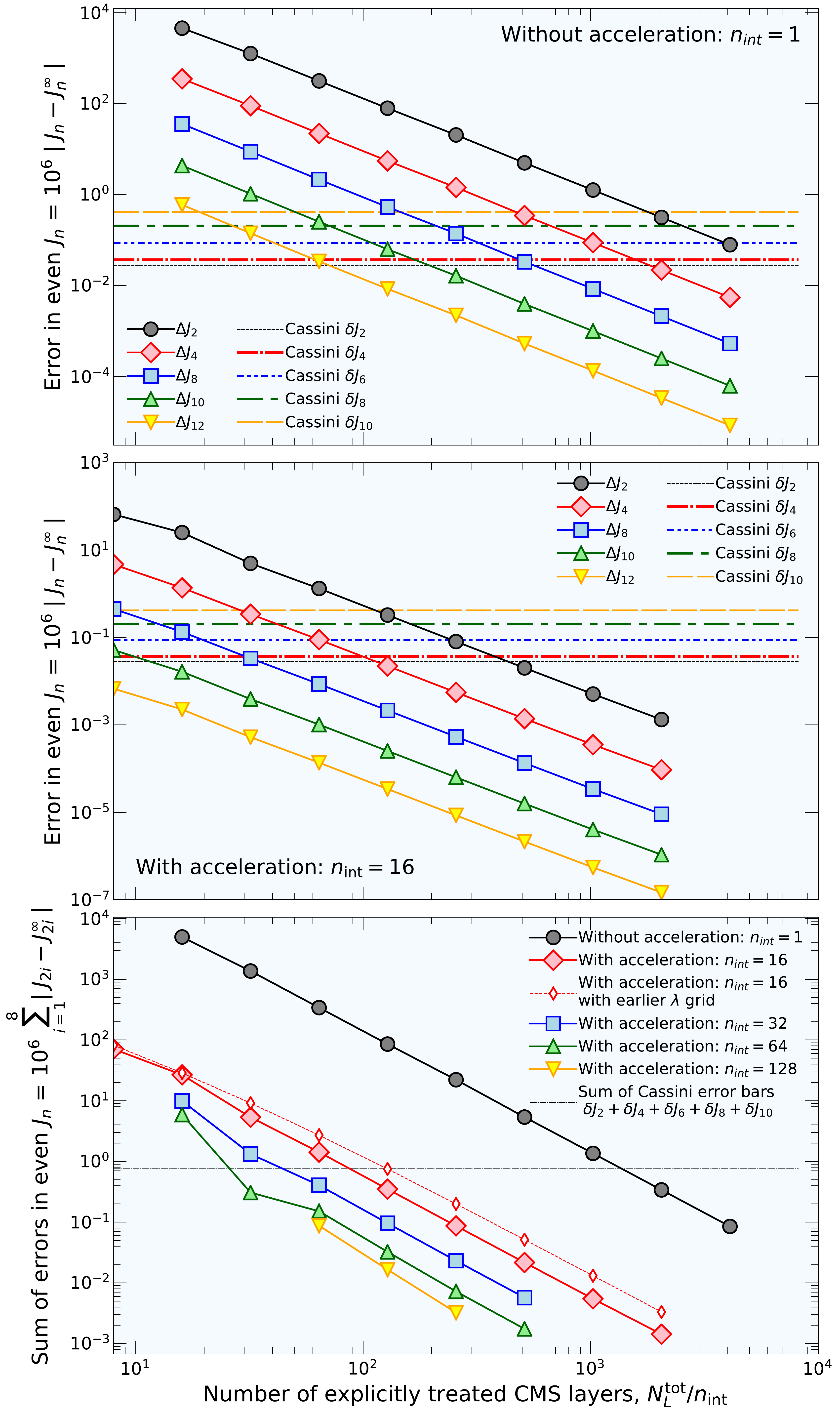}
 \caption{ Discretization error of the gravity harmonics calculated with
  the CMS method as a function of the number of spheroids. The
  horizontal lines show the 1-$\sigma$ uncertainties of the {\it Cassini}
  measurements of the even gravity harmonics, $J_n$. The top panels
  show how the errors decay with increasing number of layers
  for calculations without acceleration. The mid panel
  displays results obtained with the accelerated CMS method where only
  one in $n_{\rm int}=16$ layers are treated explicitly. 512 explicit
  layers (total of 8192) are sufficient to reduced the error in all
  calculated gravity harmonics below the uncertainties of the
  observations. Without the acceleration,
  well over 4000 layers is required for this level of precision,
  as the top panel shows. The bottom panel
  compares results derived with different acceleration factors,
  $n_{\rm int}$.  For $n_{\rm int}=16$, the effects of two different
  $\lambda$ discretization schemes are compared.
  \label{fig1} }
\end{figure}

Having access to extrapolated values, $J_n(\infty)$, allows us to
study how the discretization error decays with increasing $N_L$ and
to evaluate the effectiveness of the acceleration scheme. All curves
in Fig.~\ref{fig1} show that the discretization error decays
quadratically as $N_L^{-2}$. The top panel shows the behavior of the
original method before any acceleration was introduced. For $J_{12}$,
one finds that only 32 layers are needed for the discretization error
to be less than the error bar of the {\it Cassini} measurements because the
uncertainty is comparatively large for this gravity
coefficient. Conversely $J_2$ has been measured with a much higher
precision and even CMS calculations with 4096 layers are not
sufficient to meet the accuracy of the measurements.

The middle panel of Fig.~\ref{fig1} shows the discretization error of
accelerated CMS method with acceleration factor, $n_{int}=16$. The
results show that calculations with 512 explicit layers
($N_L^{\rm tot}=16384$) are sufficiently accurate to reduce the
discretization error of computed gravity coefficients below the uncertainty level of
the {\it Cassini}
measurements. This demonstrates that with the acceleration technique
is very effective and enables us to match the accuracy of {\it Juno} and
{\it Cassini} measurements within the CMS framework. 

In the lower panel of Fig.~\ref{fig1}, the discretization error of
different $J_n$ have been combined in order to compare results for
different acceleration factors, $n_{\rm int}$. The figure confirms
that an increase in $n_{\rm int}$ leads to a significant reduction of
the discretization error when results are compared for the same 
number layers that are treated explicitly, $N_L^{tot}/n_{\rm int}$,
which is also a measure of the computational cost.

\begin{figure}[ht!]
\includegraphics[width=8cm]{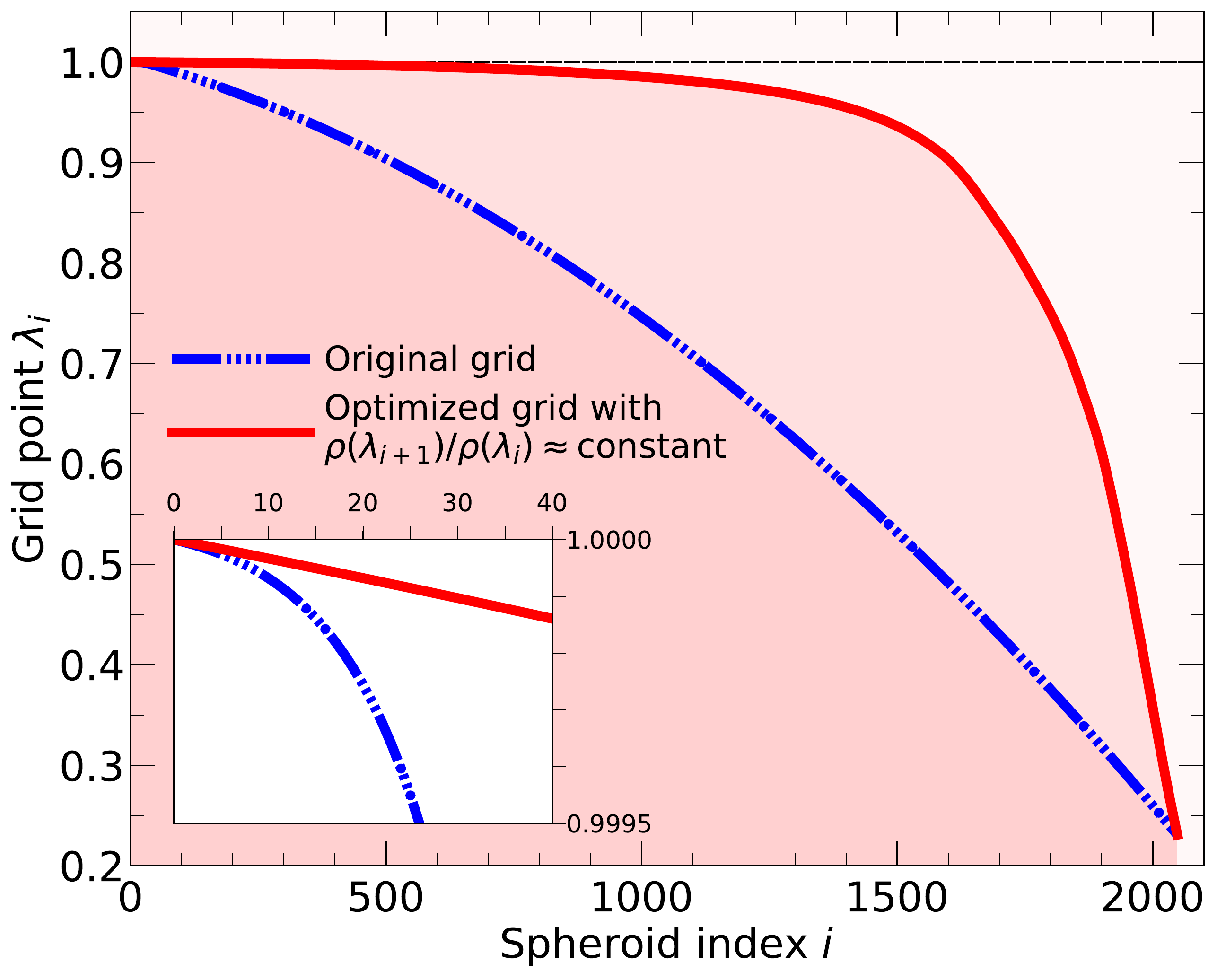}
\caption{Comparison of our original and optimized $\lambda$ grids for
  an interior model with 2049 spheroids and $r_C$=0.231.}
\label{lambda_grid}
\end{figure}

The lower panel of Fig.~\ref{fig1} also compares the discretization
error that arises from two different $\lambda$ grids. The choice of
$\lambda$ grid has an impact on how many spheroids are needed to reach
a certain level of accuracy. We show results derived with an earlier
$\lambda$ grid from \citet{WHM16a}, which was constructed by employing
a denser mesh of spheroids in the atmosphere and outer layers of the
planet where the density changes the most. We then developed an
alternate approach with the aim of constructing an optimal $\lambda$
grid that further reduces the discretization error. This error arises
from contrast in density between two adjacent spheroids. To minimize
this error, we construct a $\lambda$ grid such that the relative
difference in density is the same for all pairs of adjacent spheroids
throughout the planet. This automatically places more layers in the
atmosphere, where the density changes most rapidly. We construct our
optimized $\lambda$ grid by starting from a converged CMS calculation
with our original grid, which provides us with series of
$\rho(\lambda_i)$ points that we can interpolate. We construct a
geometric grid of $\rho_i$ values that the spans the interval between
the lowest and highest density in our model while keeping
$\rho_{i+1} / \rho_i $ constant. We derive our optimized $\lambda_i$
grid by solving $\rho(\lambda_i)=\rho_i$. In Fig.~\ref{lambda_grid}
and Tab.~\ref{lambda_table}, we compare the original and optimized
$\lambda$ grids. During the optimization, more grid points are placed
in outer region of the planet where the density changes most
rapidly. However, the inset of Fig.~\ref{lambda_grid} shows that the
slopes of the two grid functions is very similar near $\lambda=0$. In
this region the grid space should be a fraction of the scale height of
the atmosphere.

In limit of $N_l \to \infty$, CMS calculations with both $\lambda$
grids will converge to identical results because the discretization
errors will gradually dimish in every part of the interior. However,
an optimized $\lambda$ grid may approach this limit more rapidly. The
lower panel of Fig.~\ref{fig1} shows that our optimized $\lambda$ grid
reduces the discretization error by a factor of 2.3 when compared to
our original grid for the same number of spheroids. For this reason,
we employ the optimized grid in all following calculations.

\begin{table}
  \caption{ Original and optimized $\lambda$ grids for
  an interior model with 2049 sphoeroids and $r_C$=0.231. The entire table will be published online.\label{lambda_table}}
\begin{tabular}{|c|r|r|}
\hline
  Spheroid index $i$ & Original $\lambda_i$ & Optimized $\lambda_i$ \\
\hline
0 & 1.0000000000 &   1.0000000000\\
1 & 0.9999958561 &   0.9999966866\\
2 & 0.9999912702 &   0.9999933632\\
3 & 0.9999861950 &   0.9999900322\\
$\ldots$  &  $\ldots$            & $\ldots$\\
2047 & 0.2316114866 & 0.2338981435\\
2048 & 0.2310000000 & 0.2310000000\\
\hline
\end{tabular}
\end{table}

\subsection{Planet models with polytrope index 1}

Here we revisit standard planetary interior models that approximate
the equation of state throughout the interior by a polytropic equation
of state, $P(\rho) = K \rho^{1+1/n}$ with index $n=1$. The constant
$K$ is adjusted so that planet's total mass equals 1. Under these
assumptions, potential and density are proportional and the planet's
surface is given by $P=\rho=0$. \citet{wisdom2016} studied the
properties of such planet models in great detail and compared the
predictions from the consistent level curve (CLC) technique and from
the CMS method. Here we present a comparison with our accelerated CMS
approach, which allows us to control density discretization error more
carefully. We benchmark our results against \citet{wisdom2016} using
the identical value of $q_{\rm rot}$=0.089195487.

\begin{figure}[ht!]
\includegraphics[width=8cm]{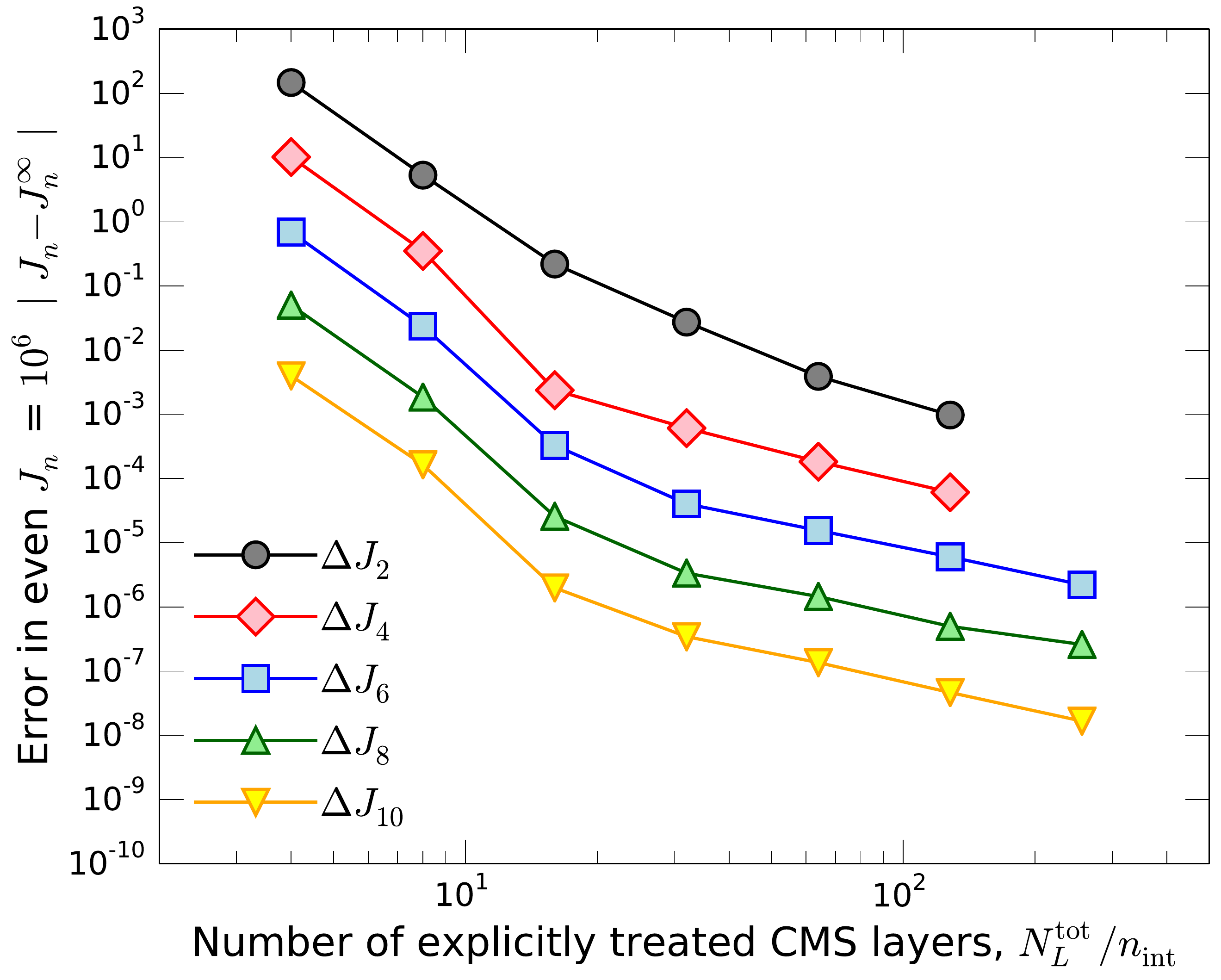}
\caption{Discretization error in the gravity harmonics of polytrope
  index 1 planet models. The error of all gravity harmonics decays
  with increasing spheroid number, as we have seen for the Saturn
  interior models in Fig.~\ref{fig1}. All calculations for this figure
  were performed with $n_{\rm int}=256$.}
\label{polytrope}
\end{figure}

In Fig.~\ref{polytrope}, we show how discretization errors decay with
increasing number of spheroids. Overall the behavior is similar to
that of our more realistic Saturn interior model in Fig.~\ref{fig1}.

We choose a acceleration factor of $n_{\rm int}=256$ and performed a
set of polytrope index 1 model calculations with increasing precision.
The number of explicitly treated layers, $N_L / n_{\rm int}$ were varied
between $2^2$ and $2^9$, which brought up the total number of layers
to 131072 in our largest calculations, which is an increase of three
orders of magnitude compared to earlier CMS calculations. We analyze
how our results improved with increasing layer number and report the
converged digits in Tab.~\ref{tab_poly}. The agreement with the CLC
predictions is excellent. All coefficients $J_2$ through $J_{20}$
agree to 6, 7, or 8 significant digits, which is a better agreement
than was reported in \citet{wisdom2016} where predictions from the CLC
approach and the non-accelerated CMS method were compared.

\begin{table}
  \caption{ Gravity coefficients for the polytrope index 1 planet models derived
    with the accelerated CMS (this work) and CLC~\citep{wisdom2016} methods. \label{tab_poly}}
\begin{tabular}{lrr}
Gravity coefficient & CMS & CLC\\
$10^{2}  \times J_{2}$  &    1.3988511  &  1.398851090\\
$10^{4}  \times J_{4}$  & $-$5.318281~$\,$ & $-$5.318281001093 \\
$10^{5}  \times J_{6}$  &    3.0118324  &  3.011832290534\\
$10^{6}  \times J_{8}$  & $-$2.1321158  & $-$2.132115710725 \\
$10^{7}  \times J_{10}$ &    1.74067138 &  1.740671195866\\
$10^{8}  \times J_{12}$ & $-$1.56821961 & $-$1.568219505563\\
$10^{9}  \times J_{14}$ &    1.51809944 &  1.518099226841\\
$10^{10} \times J_{16}$ & $-$1.5519853 & $-$1.551985081630\\
$10^{11} \times J_{18}$ &    1.6559259 & 1.655925984019\\
$10^{12} \times J_{20}$ & $-$1.8285783 & $-$1.828574676495 \\
\end{tabular}
\end{table}

\subsection{ Parameter Optimization }

The primary goal of the model optimization is the generation of Saturn
interior models that reproduce the observed gravity harmonics. The
agreement between models and observations is typically expressed in
some form of a $\chi^2$ function. Here we use,
\begin{equation}
  \chi^2_J = \sum\limits_{i=1}^5 \left[ \frac{ J_{2i}^{\rm observed} - J_{2i}^{\rm model} }{ \delta J_{2i}^{\rm observed} } \right]^2
  \quad,
  \label{chi_J}
\end{equation}  
where $ \delta J_{2i}^{\rm observed}$ are the 1-$\sigma$ uncertainties
in the observations. Typically $J_2$ is measured with much higher
precision than the higher order harmonics. To deal with this
imbalance, we find solutions that satisfy
$J_{2i}^{\rm observed} = J_{2i}^{\rm model}$ exactly by adjusting one
model parameter like $Z_{\rm mol}$ or $Z_{\rm met}$ before $\chi^2_J$
is evaluated. This optimization is performed for converged CMS
models that have reached hydrostatic balance and have matched 
the planet's
total mass by adjusting the core mass.

While Eq.~\ref{chi_J} is certainly the most important optimization
criterion, there are a number of other well motivated constraints to
consider. For example, one would want to guide to the parameter
optimization towards models with pressures $P_1$ and $P_2$ are close
to the assumed immiscibility curve in Fig.~\ref{fig_imm}. From the
assumed molecular and metallic adiabats, we can infer the temperatures
$T_1$ and $T_2$ that correspond to both pressures. For both pairs
$P_1$-$T_1$ and $P_2$-$T_2$, we find the closest points on the
immiscibility curve, $P^*_1$-$T^*_1$ and $P^*_2$-$T^*_2$, that
minimize the following immiscibility penalty function,
\begin{equation}
  \chi^2_{\rm H-He} = \sum\limits_{i=1}^{2}C_P \left| \frac{P^*_i-P_i}{P_i}  \right| + C_T \left| \frac{T^*_i-T_i}{T_i}  \right| 
  \quad,
  \label{chi_H-He}
\end{equation}  
before we add the resulting minimum value to the total $\chi^2$. $C_P$
and $C_T$ are weights that must be balanced with those in other
$\chi^2$ terms. We set $C_T/C_P=2$. We chose not to square the
individual terms in Eq.~\ref{chi_H-He} because, without an
experimental confirmation of our immiscibility curve, we do not want
large deviations to enter quadratically. In Fig.~\ref{fig_imm}, we
shows some representative models to illustrate how much variation is
in the $P_1$-$T_1$ and $P_2$-$T_2$ in our ensemble of
models. Implicitly the $\chi^2_{\rm H-He}$ term also introduces a penalty
for metallic adiabats that are too hot to be compatible with the
assumed immiscibility curve. 

Upon first introducing differential rotation into our CMS models, we
realized that a super-rotating equatorial jet improved the match to
observed gravity harmonics considerably. Furthermore, for $l \ge 0.8$,
the inferred rotation profile was compatible with the wind speeds that
were derived from tracking the clouds in Saturn's
atmosphere~\citep{SL2000,GM2011}. From this point on, we favored models
that matched those observations by introducing the following cloud
penalty function,
\begin{equation}
  \chi^2_{\rm clouds} = C_{\rm clouds} \sum_{k ~{\rm with}~ l_k>0.8 }    \left| \omega_{\rm observed}(l_k) - \omega_{\rm model}(l_k)  \right|
  \quad,
  \label{chi_clouds}
\end{equation}
where we sum over the knots, $l_k$, in the outer region of our
rotation profile that often lead to good agreement with the cloud
tracking observations. $C_{\rm clouds}$ plays the role of a weight.

Finally we introduce one more penalty function,
\begin{equation}
  \chi^2_{\rm curvature} = C_{\rm curvature} \sum_{{\rm all}~k}    \left[ \omega''_{\rm model}(l_k)  \right]^2
  \quad,
  \label{chi_curvature}
\end{equation}
that favors smooth rotation profiles by penalizing large values in the
second derivative of our rotational profile. 

\begin{figure}[ht!]
\includegraphics[width=8cm]{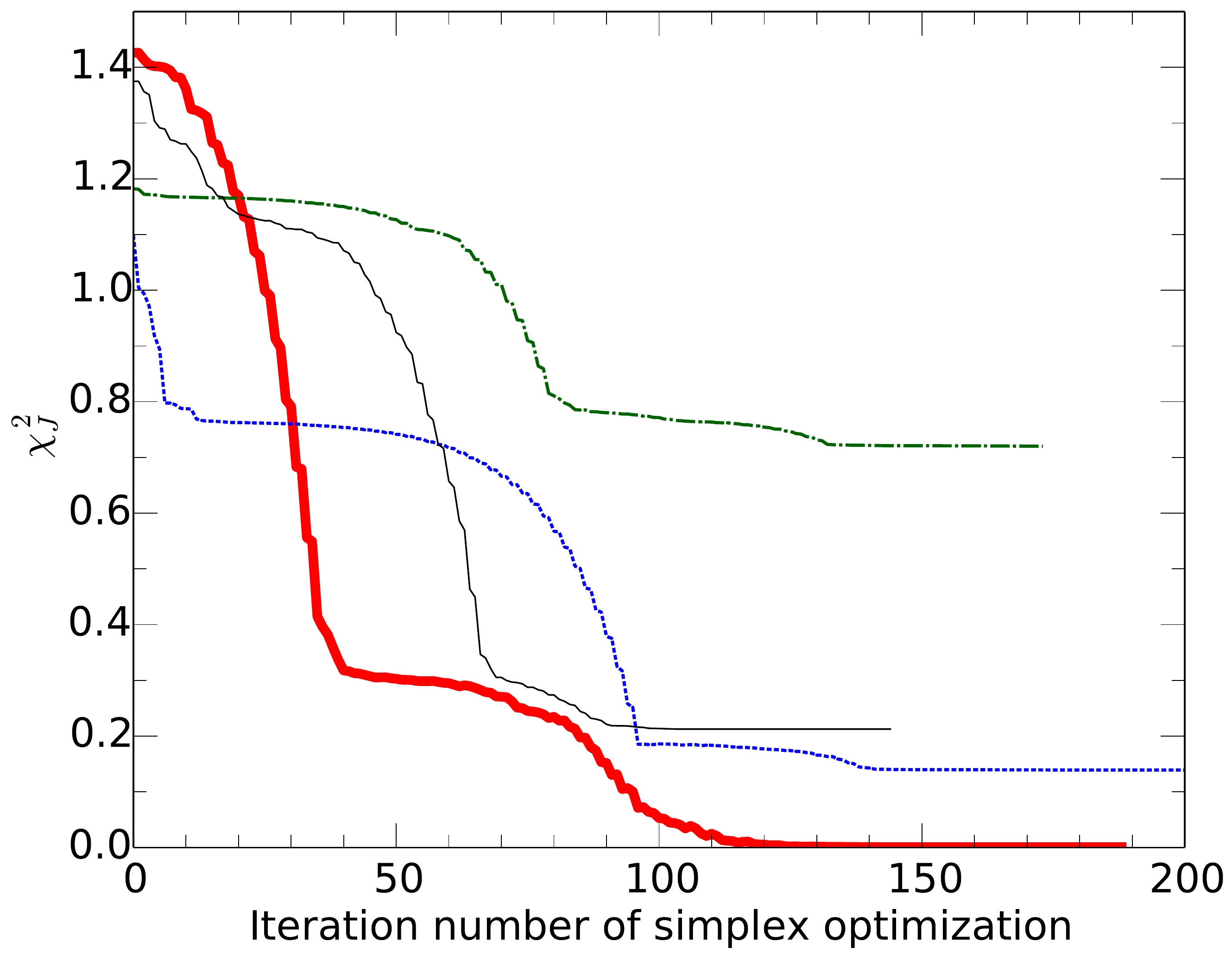}
  \caption{$\chi_J^2$ deviations (see Eq.~\ref{chi_J}) between the calculated and the observed
  gravity harmonics during the model optimization with the simplex
  algorithm. Only one model optimization (thick solid line) succeeded in
  converging to a state that matched the spacecraft measurements
  well. \label{fig_chi2}}
\end{figure}

We assume $Y_{\rm met} \ge Y_{\rm mol}$ and
$Z_{\rm met} \ge Z_{\rm mol}$, because we assume that helium rain can
only lead to an enrichment of the metallic layer in helium and in
heavy elements~\citep{WilsonMilitzer2010}. We also constrain the helium
abundance of the entire envelope to match solar proportions.

We add Eqs.~\ref{chi_J} through \ref{chi_curvature} to obtain one total $\chi^2$
function that we employ to optimize the model parameters in Tab.~\ref{tab3}. This
turns out to be a very challenging optimization problem, because many parameters are
strongly coupled and some optimization criteria are interdependent. We use the
simplex algorithm~\citep{numerical_recipes} for the optimization since it does not
require any derivatives of $\chi^2$ with respect to the optimization parameters,
which are not available in analytical form. With this algorithm, it was very
challenging to generate models that matched observed gravity data. In many cases, the
algorithm gets stuck in a local minimum\footnote{We computed the derivative
    numerically and tested the BFGS~\citep{BFGS} optimization algorithm but this did
not lead to an algorithm that is more efficient overall because of the cost of
computing the derivative with finite differences.}.  Fig.~\ref{fig_chi2} shows a
couple of examples of the $\chi_J^2$ evolution during the simplex
optimization. However, in 17 independent cases, the
optimization succeeded and we were able to match the gravity harmonics within the
uncertainties of the observations. We subsequently used these 17 solutions as
starting points for Markov chain Monte Carlo calculations in order to map out the
allowed parameter regions. We confirmed that all 17 original solutions belong to the
same parameter region and one can go smoothly from one to the other. This provides
strong evidence that the entire solution space is connected.

\subsection{Effects of an upper atmosphere}

All CMS calculations presented so far, start from an outermost spheroid
with the pressure of 0.1 bar that was anchored at the equatorial
radius $a$. We had thereby neglected the effects of the tenuous upper
atmosphere that extends from the 0.1 bar level out into space. To
study the effects of this upper atmosphere quantitatively, we added
64, 128, and 256 outer spheroids to CMS calculations with 512, 1024,
and 2048 layers, respectively. The number of the additional spheroids
was chosen such that range of pressure extended down to at least 1
mbar. The original outer spheroid is still associated with a pressure
of 0.1 bar and remains anchored at the equatorial radius $a$. For all
spheroids interior to this spheroid, we update the pressure according to
Eq.~\ref{P_update} as we did before. However, for all the additional
 exterior spheroids, we update the pressure with decreasing spheroid index according to,
\begin{equation}
    P_{i}^{\rm (new)} = P_{i+1}^{\rm (new)} + \rho_{i} \, (U_{i+1}-U_{i}).
    \label{P_update2}
\end{equation}
For simplicity, we assume an isothermal upper atmosphere with a
temperature set to the value at 0.1 bar. In all other respects the
additional exterior spheroids are treated in the same way as the
interior spheroids. In principle, the temperature treatment could be made more
realistic but, as we will show, the effects on the computed
gravitational moments are negligible because there is so little mass
outside of the 0.1 bar level.

\begin{table}
  \caption{ Correction to the computed gravitional moments due to upper atmosphere that extends from 100 to 0.9 mbar.\label{tab_delta_J}}
\begin{tabular}{lr}

$n$ & $10^6 \times \Delta J_n$ \\
\hline
2 & $+6.3 \times 10^{-3}$\\
4 & $-1.1 \times 10^{-3}$\\
6 & $+2.5 \times 10^{-4}$\\
8 & $-7.6 \times 10^{-5}$\\
10& $+2.2 \times 10^{-5}$\\
12& $-1.7 \times 10^{-6}$\\
14& $-2.3 \times 10^{-6}$\\
16& $+4.3 \times 10^{-7}$
\end{tabular}
\end{table}

Our extended CMS calculations converged to the same level of accuracy as
they had previously. The pressure of the new outermost spheroid converged to 0.9
mbar. The fractional mass outside of 0.1 bar level was found to be
only $ 7.5 \times 10^{-8}$. This mass correction can also be
interpreted as a change to the gravity coefficient $J_0$ (see
Eq.~\ref{standard_J}), which helps one to gauge the magnitude of the
correction to the other gravity coefficients. In
table~\ref{tab_delta_J}, we provide the differences in the
gravitational moments between our CMS calculation that included an
extended atmosphere to 0.9 mbar and our original calculations that
terminate at 0.1 bar. All values decay smoothly with increasing order
$n$. For all the gravity coefficients that have been determined by the
{\it Cassini} spacecraft, $J_2$ through $J_{10}$, one finds that the
correction due to the upper atmosphere is at least 18 times smaller
than the uncertainties of the {\it Cassini}
measurements~\citep{Iess2019}. For this reason, we conclude that our
standard CMS calculations starting from the 0.1 bar level are
sufficiently accurate for this study.

\section{Results} \label{sec:results} 

\subsection{Predictions for interior parameters} \label{sec:predict}

In Fig.~\ref{DR}, we plot the rotational profiles that have emerged from our Monte
Carlo calculations. Two prominent features are common to all models. There is a
super-rotating equatorial jet in the equatorial region that rotates up to 4\% faster
than the deep interior. This behavior is in agreement with the observed wind speeds
from tracking the cloud motion in Saturn's visible atmosphere~\citep{SL2000,GM2011} and
we have thus favored the sampling of such models by introducing the term $\chi^2_{\rm
clouds}$ in Eq.~\ref{chi_clouds}. At a distance of approximately 50$\,$000 km from
the axis of rotation, our models require a sub-rotating region with a
flow about 1\% slower than in the deep interior. This feature is not 
observed in the cloud motion at the surface, but is a common feature to all of our
models that match the {\it Cassini} gravity harmonics. Both the super-rotating
equatoral jet, and the sub-rotating feature are present regardless of the
value we 
assume for the rotation period of the deep anterior.

In Fig.~\ref{fig_para}, we compare the predictions from ensembles of
models that we generated with MC sampling for a range of core radii
and rotation periods for the deep interior. In panel (a), we plot the
amount of heavy elements in the envelope against the core mass. When one
compares models for the same core radius of $r_C=0.2$, a simple trend
emerges. With increasing rotation period, the amount of heavy elements in the
atmosphere decreases from approximately 4-fold to 1.2-fold the solar
value ($Z_{\rm solar} = 0.0153$ \citet{Lodders2010}) while the core
mass increases from approximately 15.3 to 16.9 Earth masses. Larger
variations in the predicted core masses are seen when the fractional
core radius is varied between 0.188 (rocky composition) and 0.231
(rock-ice core in Callisto's proportion \citep{Kuskov2005}. A smaller core radius
leads to a smaller core masses because the H-He mixture that surrounds
that core is exposed to higher pressure, which increases its density
and lets it mimic the behavior of the dense core to a larger extent
than this is the case in models with larger core radii. Therefore, the
uncertainty of the core composition is the primary reason why the
predicted core masses vary between 15 and 18 Earth masses.

In Fig.~\ref{fig_para}b, we plot the combined enrichment in helium and heavy
elements in the metallic layer against the entropy in this layer. Since a higher entropy implies
a higher temperature and thus a slightly lower density, the enrichment rises
with increasing entropy. We find the models with a very long rotation period
of 10:45:45 h and 10:47:06 h are confined to a very narrow region of available
parameter space predicting the lowest enrichment and the highest entropy for
the metallic layer. The long period models appear similarly confined in
Fig.~\ref{fig_para}c where they predict almost no helium rain had occurred
while models with shorter rotation periods predict various amount of helium
rain. $Y_{\rm mol}$ values as low as 0.19 are included.  $Y_{\rm mol}$ and
$Y_{\rm met}$ are tightly correlated in this figure because we assume the
envelope overall has a solar helium abundance.  

In Fig.~\ref{fig_para}d, we compare the heavy element abundances in the
molecular and metallic layers. Within the model constraint of $Z_{\rm met} \ge
Z_{\rm mol}$, a wide range of super-solar enrichments are predicted by our
ensembles of models. There are plenty of models with $Z_{\rm met} \approx
Z_{\rm mol}$, which is in contrast to recent Jupiter models that required a
different amounts of enrichments in the two layers~\citep{Wahl2017a}. 
In Fig.~\ref{fig_para}e
and f, we plot the heavy element against the helium abundances in the molecular
and metallic layers, respectively. While both quantities are strongly
correlated in the molecular layer, there appears to be much more flexibility in
the metallic layer. One reason for this is that a range of $S_{\rm met}$ values are
permitted in our models while the entropy in the molecular layer is tied to
the temperature at the 1 bar level. In Fig.~\ref{fig_para}e, one can identify
a consistent trend for models with longer rotation periods to predict larger
values $Y_{\rm mol}+Z_{\rm mol}$ and thus a slightly higher density for the
molecular layer.

The models with shorter rotation periods produce $Y_{\rm mol}$ that
are compatible with reanalyzed Voyager measurements of atmospheric
helium, $\sim$0.6--0.8$\times$solar \citep{Conrath2000}, while the
models with longer rotation rates require less depletion of helium in
the outer envelope.  Observational constraints on $Z_{\rm mol}$ are
uncertain; \citet{Fletcher2009} observed atmospheric methane
concentrations consistent with $\sim$9$\times$solar enrichment of
carbon. There are no direct measurements of the abundance of
atmospheric oxygen, the heavy element with the most significant
contribution to the density and by consequence the gravity. Other
heavy element ratios observations include both much lower (N/H
$\sim$3$\times$ solar) and higher S/H $\sim$13$\times$ solar),
although these differences might reflect model dependence in
determining the bulk elemental abundance from, or from measuring
regions of the atmosphere that our not well mixed~\citep{Atreya2019}.

The models presented here predict values of both $Z_{\rm mol}$ and
$Z_{met}$ between 1--4$\times$solar for a uniform enrichment of all
heavy elements, which is lower than the observed enrichment in carbon.
It is worth noting that in-situ measurements of Jupiter's atmosphere
up to 22 bars by the Galileo entry probe showed significant depletion
in oxygen compared to carbon, but it is an outstanding question
whether this accurately reflects the overall composition of Jupiter's
molecular envelope. The heavy element content predicted by the models
are also sensitive to temperature of the adiabat.  For Saturn,
atmospheric temperature has never been measured {\it in situ}. So if
$S_{\rm mol}$ is higher than we expect, this tradeoff could account
for higher concentrations of heavy elements, without significantly
affecting the other model predictions.

\subsection{Oblateness and Rotation Period} \label{sec:Ob}

\begin{figure}[ht!]
\includegraphics[width=8cm]{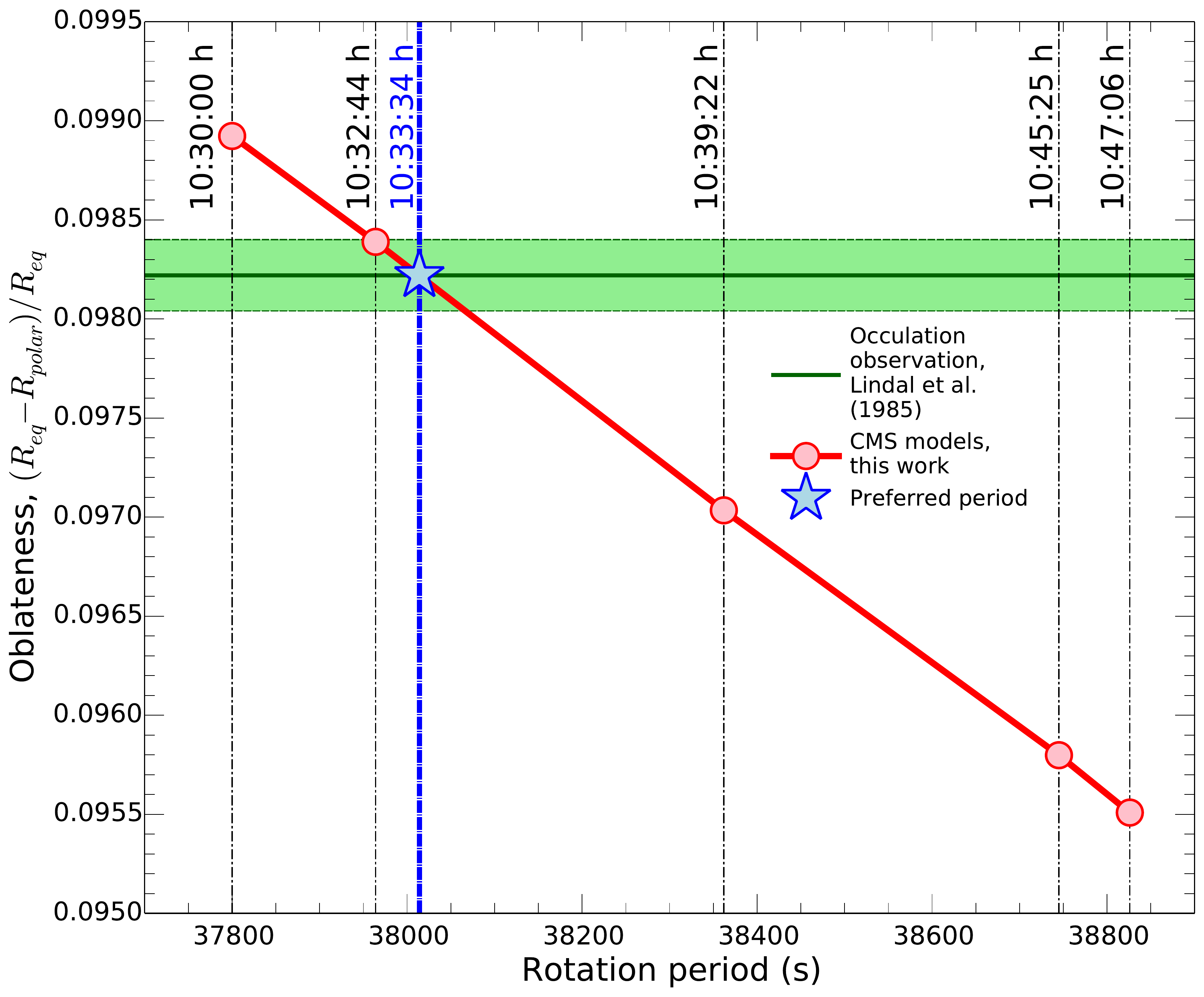}
  \caption{Oblateness derived from CMS models with different rotation
      periods compared to the radio occultation oblateness measurement by
      \citet{Lindal1985}, which determined an oblateness of 0.09822 $\pm$
    0.00018. Based on this comparison, we favor an rotation period of
    10:33:34 h $\pm$ 55 s.
    \label{fig_period}
  }
\end{figure}

While the rotation period of Jupiter's interior has been determined
with high accuracy from magnetic field observations, the rotation
period of Saturn's deep interior remains uncertain due to the
remarkable alignment between the dipole field and the axis of
rotation. However, the rotation period used in CMS calculations of a
planet significantly
affects its shape. Saturn's oblateness, $(R_{eq}-R_{polar})/R_{eq}$,
has been measured with radio occultation measurements of the  {\it Pioneer} and
{\it Voyager}
spacecraft~\citep{Lindal1985}. Anderson and
Schubert~\citep{Anderson2007} constructed interior models with uniform
rotation that matched observed oblateness and pre-{\it Cassini} gravity
coefficients $J_2$, $J_4$, and $J_6$.  They derived a rotation period
of 10:32:44 h, which is significantly shorter than the system III
period of 10:39:22 h~\citep{desch1981}, as well as {\it Cassini} predictions of
10:45:45 h and 10:47:06 h \citep{giampieri2006,Helled2015}.

In Fig.~\ref{fig_period}, we compare models with differential rotation
that we constructed for five different rotation periods ranging from
10:30:00 to 10:47:06 h. For all periods, it is possible to construct
interior models with differential rotation that match all even gravity
coefficients. However, the oblateness sensitively depends on rotation
period that is assumed for the planet's deep interior. In
Fig.~\ref{fig_period}, we compare the oblateness that derived from our models with
the radio occultation measurements by the {\it Pioneer} and
{\it Voyager}
spacecraft~\citep{Lindal1985}. We find that rotation periods of
10:33:34 h $\pm$ 55 s are consistent with these observations, which
represents a modest 1-$\sigma$ increase over the earlier determination
of 10:32:44 h by Anderson and Schubert~\citep{Anderson2007} who
performed a similar analysis based on interior models with uniform
rotation. Thus, the 50 s difference can primarily be attributed to
effects of differential rotation.

Our determination of Saturn's rotation rate is in remarkably good
agreement with the value of 10:33:38$\,$h$^{+112\rm s}_{~-89 \rm s}$
inferred from waves observed in Saturn's rings \citep{Mankovich2019},
even though the interior models for this analysis were
constructed without considering differential rotation.

In Fig.~\ref{fig_para2}, we compare predictions from models with our
preferred rotation period of 10:33:34 h to those based on the
system III rotation period of 10:39:22 h. We still predict a core mass
range from 15 to 18 Earth masses, primarily set by the
uncertainty in the core composition. When we compare
Figs. ~\ref{fig_para}d and ~\ref{fig_para2}b, we find the range of
$Z_{\rm met}$ is considerably narrowed if the rotation period is set
to 10:33:34 h. Most $Z_{\rm met}$ values now fall between values
between 1.8 and 2.5 $Z_{\rm solar}$ while in Fig.~\ref{fig_para}d, the
smaller and larger $Z_{\rm met}$ values came from models with longer
and shorter rotation periods, now disfavored because of the
oblateness constraint. The $Z_{\rm mol}$ values vary between 1 and
3-fold $Z_{\rm solar}$ as before.

Finally, in Fig.~\ref{fig_para2}c, $Z_{\rm mol}$ and $Y_{\rm mol}$ are
now fairly tightly correlated when we assume a rotation period of
10:33:34 h. These predictions can, in principle, be verified with remote
observations or by an entry probe on a future missions.

\section{Conclusions}

We have presented an accelerated version of CMS that
allows us to construct planetary interior models with many more
layers than before and also enables construction of ensembles of
models using Monte Carlo methods to efficiently optimize the
parameters of individual models. We have applied this accelerated CMS
method to construct models for Saturn's interiors with differential
rotation on cylinders, which permitted us to match the unexpectedly
large values of the gravity harmonics $J_6$, $J_8$, and $J_{10}$ that
the {\it Cassini} spacecraft measured during its Grand Finale orbits around
Saturn. From our interior models we infer that Saturn has a massive
core of $\sim$15--18 Earth masses and there are additional heavy
elements worth 1.5--5 Earth masses distributed throughout its
envelope. In our models, we have also varied the rotation period of
Saturn's deep interior and studied the effects on Saturn's 
oblateness. By matching occultation measurements of  spacecraft
we predict a rotation period of $10:33:34$h $\pm$ 55s for Saturn's
deep interior.


\acknowledgments{The authors acknowledge support from the NASA
  missions {\it Cassini} and {\it Juno}. In part, the University of California
  supported this work through multicampus research award 00013725 for
  the Center for Frontiers in High Energy Density Science.}



\widetext

\begin{table}
\caption{ Parameters and constraints in our Saturn models 
\label{tab3}}
\begin{tabular}{ll}
\hline
$S_{\rm mol}$ & Entropy H-He gas throughout the molecular layer. \\
             & Constrained to match Saturn’s 1 bar temperature of 142.6 K
             \citep{Lindal1981}\\
             $Y_{\rm mol}$ & Helium mass fraction in the molecular layer. Constraint: $
             Y_{\rm mol} \le Y_{\rm
solar}=0.2741$~\citep{Lodders03}.\\
$Z_{\rm mol}$ & Mass fraction of heavy Z elements in the molecular layer.\\
$S_{\rm met}$ & Entropy H-He gas throughout the metallic layer. $S_{\rm met} \ge S_{\rm mol}$\\ 
$Y_{\rm met}$ & Helium mass fraction in the metallic layer. Adjusted as function of $Y_{\rm mol}$ \\
             & to keep the overall composition of the envelope at $Y_{\rm solar}$.\\
$Z_{\rm met}$ & Mass fraction of heavy Z elements in the metallic layer. Constraint: $Z_{\rm met} \ge Z_{\rm mol}$.\\
$P_1$ & Starting pressure of the helium rain layer (high pressure end of molecular layer)\\
$P_2$ & Ending pressure of the helium rain layer (low pressure beginning of metallic layer)\\
Core mass & We assumed a compact core composed of heavy elements \\
& with a sharp boundary to the metallic layer, with an equatorial radius, $r_C$.\\
$\omega(l_k)$ & Angular frequency for cylinder of radius, $l_k$. Constraint: $\omega(l_k<0.7) = \omega_0$.\\
\hline
\end{tabular}
\end{table}

\begin{table}
  \caption{ Gravity coefficients predicted without acceleration
    scheme for different number of layers, $N_L$. A representative
    Saturn interior model with uniform rotation was used for this
    convergence analysis. \label{tab1}}
\begin{tabular}{r rrrrrrrr}
\hline
  $N_L$ &$J_2$ & $J_4$ & $J_6$ & $J_8$ & $J_{10}$ & $J_{12}$ & $J_{14}$ & $J_{16}$ \\
\hline
   16 & 21058.747614 & -1308.699233 &   119.572180 &   -13.513972 &     1.750222 &    -0.249135   &   0.037999 &    -0.006108\\
   32 & 17740.199991 & -1047.173049 &    92.601757 &   -10.209559 &     1.294184 &    -0.180542   &   0.027002 &    -0.004258\\
   64 & 16789.968480 &  -978.984988 &    85.931473 &    -9.416858 &     1.186999 &    -0.164690   &   0.024500 &    -0.003843\\
  128 & 16552.657945 &  -962.339808 &    84.321088 &    -9.226705 &     1.161402 &    -0.160918   &   0.023907 &    -0.003745\\
  256 & 16493.667469 &  -958.249449 &    83.927467 &    -9.180331 &     1.155164 &    -0.159999   &   0.023762 &    -0.003721\\
  512 & 16478.115263 &  -957.156333 &    83.821482 &    -9.167812 &     1.153482 &    -0.159752   &   0.023724 &    -0.003715\\
 1024 & 16474.382140 &  -956.897102 &    83.796519 &    -9.164871 &     1.153087 &    -0.159694   &   0.023714 &    -0.003713\\
 2048 & 16473.437419 &  -956.831274 &    83.790168 &    -9.164122 &     1.152986 &    -0.159679   &   0.023712 &    -0.003713\\
 4096 & 16473.201201 &  -956.814816 &    83.788580 &    -9.163935 &     1.152961 &    -0.159675   &   0.023712 &    -0.003713\\
 8192 & 16473.142127 &  -956.810701 &    83.788183 &    -9.163888 &     1.152955 &    -0.159674   &   0.023711 &    -0.003713\\
\hline
\end{tabular}
\end{table}

\begin{table}
  \caption{ Gravity coefficients for the Saturn interior model in Tab.~\ref{tab1} predicted with acceleration factor
  $n_{\rm int}=16$. The first column denotes the number of CMS layers that were
  treated explicitly and the second specifies the total layer
  number. The last row contains the extrapolated values for $N_L^{\rm tot} \to \infty$. \label{tab2}}
\begin{tabular}{rr rrrrrrrr}
\hline
  $N_L^{\rm tot}/n_{\rm int}$ & $N_L^{\rm tot}$ &$J_2$ & $J_4$ & $J_6$ & $J_8$ & $J_{10}$ & $J_{12}$ & $J_{14}$ & $J_{16}$ \\
\hline
    8 &   128 & 16538.897354 &  -961.501994 &    84.237944 &    -9.215607 &     1.159703 &    -0.160646 &     0.023863 &    -0.003738\\
   16 &   256 & 16498.268862 &  -958.176106 &    83.922106 &    -9.180301 &     1.155204 &    -0.160008 &     0.023764 &    -0.003721\\
   32 &   512 & 16478.086394 &  -957.150893 &    83.821249 &    -9.167769 &     1.153475 &    -0.159751 &     0.023723 &    -0.003715\\
   64 &  1024 & 16474.446344 &  -956.898550 &    83.796655 &    -9.164888 &     1.153089 &    -0.159694 &     0.023714 &    -0.003713\\
  128 &  2048 & 16473.448194 &  -956.831549 &    83.790193 &    -9.164125 &     1.152986 &    -0.159679 &     0.023712 &    -0.003713\\
  256 &  4096 & 16473.202955 &  -956.814864 &    83.788584 &    -9.163936 &     1.152961 &    -0.159675 &     0.023712 &    -0.003713\\
  512 &  8192 & 16473.142447 &  -956.810710 &    83.788183 &    -9.163888 &     1.152955 &    -0.159674 &     0.023711 &    -0.003713\\
 1024 & 16384 & 16473.127416 &  -956.809674 &    83.788084 &    -9.163877 &     1.152953 &    -0.159674 &     0.023711 &    -0.003713\\
 2048 & 32768 & 16473.123665 &  -956.809415 &    83.788059 &    -9.163874 &     1.152953 &    -0.159674 &     0.023711 &    -0.003713\\
$\infty$ & $\infty$ & 16473.122342 &  -956.809322 &   83.788050 &-9.163873 &    1.152952 &   -0.159674 &    0.023711 &     -0.003713\\
\hline
\end{tabular}
\end{table}

\begin{figure}
\includegraphics[width=13cm]{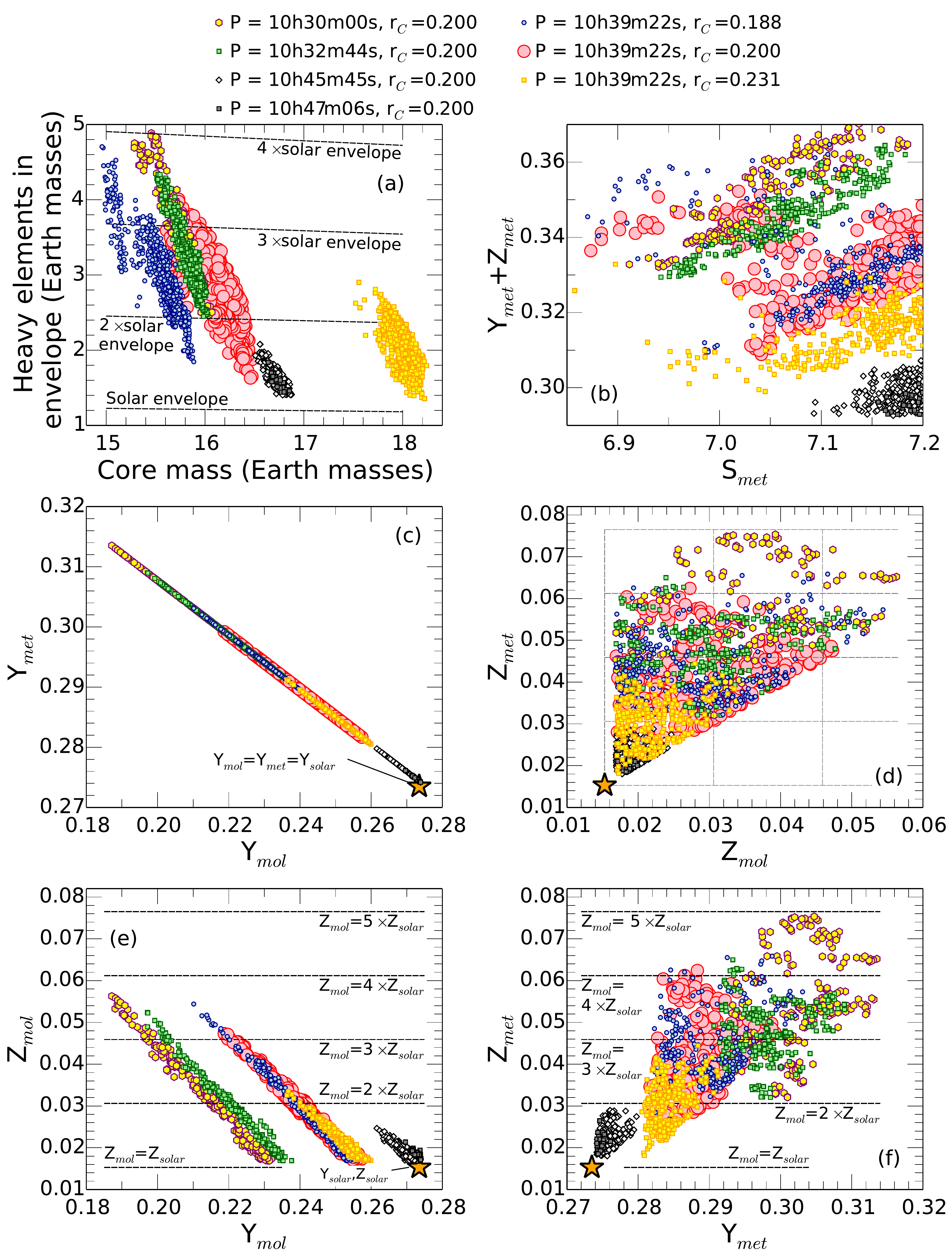}
  \caption{
    Comparison of parameters from seven sets of interior models including
      differential rotation. The assumed rotation periods
      \citep{desch1981,Gurnett2005,giampieri2006,Helled2015} and fractional core radii are
      indicated by the color and symbol, as specified in the legend. $r_{C}=0.188$
      and $r_{C}=0.231$ correspond to cores with iron-silicate and iron-silicate-ice compositions respectively.
      (a) The distribution of heavy element mass between the core and envelope.
      (b) The variation of the mass fraction elements heavier than hydrogen with
      entropy in the inner, metallic envelope.
      (c) The variation of helium mass fraction in the molecular and metallic envelopes. 
      (d) The variation of heavy element mass fraction in the molecular and metallic envelopes. 
      (e) The tradeoff between heavy element and helium mass fractions in the
      molecular envelope.
      (f) The tradeoff between heavy element and helium mass fractions in the
      metallic envelope.
      In panels (b),(c),(d),(e) and (f) solar values \citep{Lodders2010} are shown with a
      yellow star, corresponding to an assumed end-member case with no partitioning
      of helium through rain-out.
      \label{fig_para}
    }
\end{figure}

\begin{figure}
\includegraphics[width=13cm]{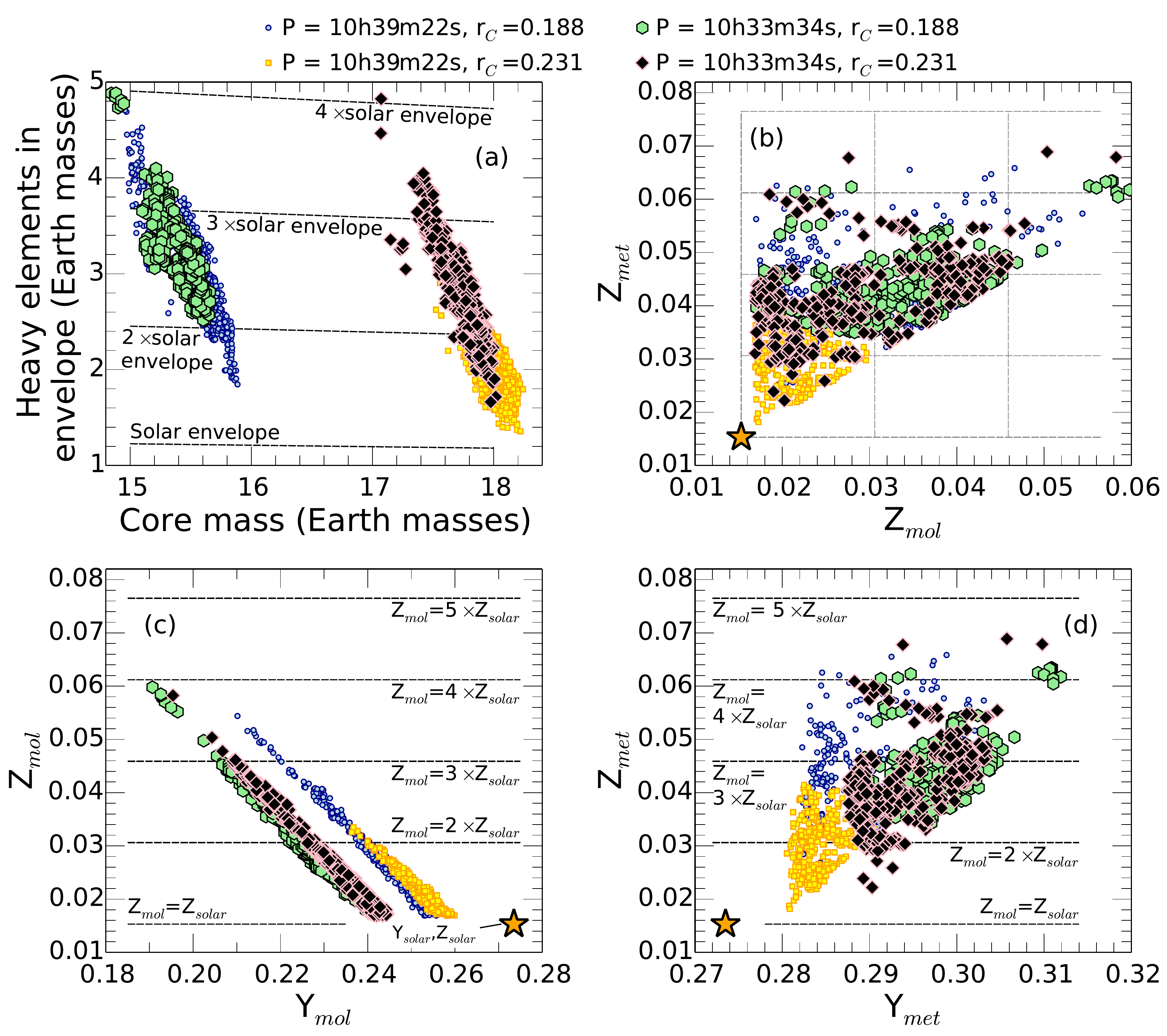}
  \caption{ Comparison of parameters from interior models with the preferred rotation
      rate of 10:33:34 h from this paper, compared to the System III
      rotation rate \citep{desch1981}.
      Core radii $r_{C}=0.188$ and $r_{C}=0.231$ correspond to rocky and rock-ice compositions respectively.
      (a) The distribution of heavy element mass between the core and envelope.
      (b) The variation of heavy element mass fraction in the molecular and metallic envelopes. 
      (c) The tradeoff between heavy element and helium mass fractions in the
      molecular envelope.
      (d) The tradeoff between heavy element and helium mass fractions in the
      metallic envelope.
      In panels (b),(c),and (d) solar values \citep{Lodders2010} are shown with a
      yellow star, corresponding solar proportions.
      \label{fig_para2}
    }
\end{figure}

\end{document}